\documentstyle[prd,eqsecnum,aps,epsf]{revtex}

\def\be{\begin{equation}}
\def\ee{\end{equation}}
\def\ba{\begin{eqnarray}}
\def\ea{\end{eqnarray}}
\newcommand{\beqn}{\begin{eqnarray}}
 \newcommand{\eeqn}{\end{eqnarray}} \newcommand{\bea}{\begin{eqnarray}} 
 \newcommand{\eea}{\end{eqnarray}}

 \newcommand{\vp}{\varphi}

\newcommand{\singlefig}[2]{
\begin{center}
\begin{minipage}{#1}
\epsfxsize=#1
\epsffile{#2}
\end{minipage}
\end{center}}
\newenvironment{figcaption}[2]{
 \vspace{0.3cm}
 \refstepcounter{figure}
 \label{#1}
 \begin{center}
 \begin{minipage}{#2}
 \begingroup \small FIG. \thefigure: }{
 \endgroup
 \end{minipage}
 \end{center}}
\def\beq{\begin{equation}}
\def\eeq{\end{equation}}
\newcommand{\gsim}{\mbox{\raisebox{-1.ex}{$\stackrel
     {\textstyle>}{\textstyle\sim}$}}}
\newcommand{\lsim}{\mbox{\raisebox{-1.ex}{$\stackrel
     {\textstyle<}{\textstyle \sim}$}}}
\newcommand{\square}{\kern1pt\vbox{\hrule height
1.2pt\hbox{\vrule width 1.2pt\hskip 3pt
   \vbox{\vskip 6pt}\hskip 3pt\vrule width 0.6pt}\hrule
height 0.6pt}\kern1pt}

\begin{document}

\draft

\title{
{\bf Correlation-consistency cartography of the double 
inflation landscape}}
\author{Shinji Tsujikawa$^{1}$, David Parkinson$^{2}$ and 
Bruce A. Bassett$^{2}$} \address{$^1$ Research Center for the Early 
Universe, 
University of Tokyo, Hongo, Bunkyo-ku, Tokyo 113-0033, Japan \\[.3em]} 
\address{$^2$ Institute of Cosmology and Gravitation, University of 
Portsmouth, Mercantile House, Portsmouth PO1 2EG, \\
United Kingdom \\[.3em]} 
\date{\today} 
\maketitle
\begin{abstract}
We show explicitly some exciting features of double-inflation: 
(i) it can often lead to strongly correlated adiabatic and 
entropy (isocurvature) power spectra.  
(ii) The two-field slow-roll consistency 
relations can be violated
when the correlation is large at Hubble crossing.  (iii) The 
spectra of 
adiabatic and entropy perturbations can be strongly scale-dependent and 
tilted toward either the red or blue.  These effects are typically due to a 
light or time-dependent entropy mass and a non-negligible angular velocity 
in field space during inflation.  They are illustrated via a 
multi-parameter numerical search for correlations in two concrete models.  
The correlation is found to be particularly strong in a supersymmetric 
scenario due to rapid growth of entropy perturbations in the tachyonic 
region separating the two inflationary stages.  Our analysis suggests that 
realistic double-inflation models will provide a rich and fruitful arena 
for the application of future cosmic data sets and new approximation 
schemes which go beyond slow-roll.
\end{abstract}
\vskip 2pc

\baselineskip = 12pt

\section{Introduction}                        

One of the radical 
developments in recent inflationary research
has been the realisation -- implicit in early work \cite{early} -- 
that inflationary predictions for the CMB and large-scale structure 
(LSS) can depend sensitively on post-inflationary, but 
pre-photon-decoupling, 
physics.
This is a departure from the single-field inflationary 
paradigm \cite{pert} that has been the backbone of 
high-energy cosmology over the past 20 years.  This rather subtle paradigm 
shift can be primarily attributed to the driving force of particle physics 
inflationary models \cite{Lyth:1998xn} which necessarily involve more than 
one dynamically important field and often lead to more than one phase of 
inflation \cite{realmulti}.

The key point about multi-field models of inflation for this paper is 
that they allow for  substantial super-Hubble entropy/isocurvature 
perturbations \cite{linde1985} (see also refs.~\cite{KS,MFB}).  This 
implies a  very interesting dynamics since, {\em at linear order}, 
entropy perturbations  source adiabatic  perturbations while the 
converse is not true in the  
large scale limit \cite{Gordon:2000hv} (though see the counter-claims
in \cite{hwang}). 
Further, these entropy modes can be partially or 
completely correlated with the adiabatic modes, and this correlation 
\footnote{This mode-mode correlation is to be contrasted with the 
time-dependent correlations of \cite{LF}.} is 
both important for the CMB and sensitive to the way in which reheating 
occurs.

Our aim in this paper is to provide the first 
exhaustive study of adiabatic-entropy correlations in 
``realistic" double inflation models. 
Given that the current CMB data actually 
favour such a correlated cocktail \cite{Amendola:2001ni} 
there exists the exciting 
possibility that upcoming data will allow us to significantly 
constrain realistic inflationary parameter spaces. 

Let us briefly recap the areas discovered so far for which 
entropy perturbations can be important.

\begin{itemize}
\item {\em Perturbations in multi-field inflationary models} 
\cite{Hwang:1991aj}-\cite{Wands:2002bn} -- 
models with two or more phases of 
inflation typically lead to some correlation due to the curvature of the 
phase curves in field space.  This correlation can be preserved or 
wiped-out depending on the precise details of reheating.

\item {\em The curvaton}\cite{curv} - an entropy perturbation can
be converted into an adiabatic perturbation with a total correlation. 

\item {\em Preheating} \cite{preheat}- the non-perturbative, resonant, 
decay   of the inflaton 
can affect standard inflationary predictions for the CMB in certain special
cases where there is an entropy perturbation on large scales that is 
resonantly  amplified at preheating. 

\end{itemize}

The possibility of correlated mixtures of adiabatic and isocurvature
perturbations is both exciting and depressing for phenomenology.  Instead 
of a single (adiabatic) power spectrum, one needs a matrix of power 
spectra 
\cite{TRD,BMT} describing the full correlation network for the complex 
cosmic cocktail of fluids.
In addition the evolution of the correlation power spectra is 
very sensitive to  the way in which particle decays occur after 
inflation.  The precise  nature 
of decay channels and widths during and after reheating can preserve or 
wash-out pre-existing correlations, introducing new arbitrary parameters 
but also opening up a new window on particle physics 
beyond the inflaton potential. Multi-field models may also lead to 
significant
levels of non-Gaussianity  in the CMB transferred from the entropy to 
adiabatic
modes \cite{nongauss}. 

There are still unresolved issues in the multi-field context. In 
particular,  the validity of the slow-roll approximation has not been 
fully explored.  Indeed,  this is one of the aims of our analysis. 
In addition, new effects occur in the case when the kinetic 
terms of the scalar fields are not canonical (e.g. nonlinear sigma 
model) and hence parametrise a curved manifold, 
as occurs in the case of scalar-tensor theories  t
\cite{Starobinsky:1994mh,Garcia-Bellido:1995fz,Starobinsky:2001xq} 
and string-inspired cosmologies \cite{Finelli:2001sr}.  

An analysis of scalar perturbations in such a general situation has
been studied \cite{Garcia-Bellido:1995fz,Mukhanov:1997fw} but only
under the assumption of the slow-roll.  Even in the single-field case
the slow-roll approximation can introduce errors in the calculation of
the CMB spectrum of up to $15\%$ \cite{slowroll} and going to higher 
order in the slow-roll parameters  may be necessary \cite{Leach:2002ar}. 
The situation in the more general case is clearly more subtle.

Recently Bartolo {\em et al.} \cite{Bartolo:2001rt} investigated the 
spectra of correlated perturbations and the modification of the standard 
consistency relation, $n_T=-2r_T$, using the slow-roll analysis in the 
multi-field context (Here $n_T$ is the spectral index of the gravitational 
wave and $r_T$ is the relative amplitude of tensor to scalar 
perturbations).  According to their results, the single-field
consistency relation is significantly modified when the correlation $r_C$ 
between adiabatic and isocurvature perturbations is strong, i.e.,

\begin{itemize}
\item \underline{The first consistency relation:}
\end{itemize}
\beqn \label{consistency1a} 
r_T=-\frac{n_T}{2}\left(1-r_C^2 \right) \,. 
\eeqn 

In addition to the standard slow-roll approximation where the second-order
derivatives of scalar fields are neglected, Bartolo {\em et al.}   
assumed that the adiabatic/entropy mass and the scalar field velocity 
angle evolve slowly during the multiple phases of inflation.  While the 
latter approximation is generally valid in the single-field context, this 
is not so in models with two stages of inflation because the masses of 
field perturbations as well as the slow-roll parameters already get large 
around the end of the {\it first} stage of inflation.  Making use of 
this approximation, Bartolo {\em et al.} derived a second consistency relation 
\cite{Bartolo:2001rt} 

\begin{itemize}
\item \underline{The second consistency relation:}
\end{itemize}

\beqn \label{consistency2a} 
\left(n_C-n_S\right)r_T= -\frac{n_T}{4}\left(2n_C-n_{\cal R}-n_S 
\right)\,,
\eeqn 
where $n_{\cal R}$, $n_S$ and $n_C$ are the spectral indices
of curvature perturbations, isocurvature perturbaions and their 
correlations, respcetively.

More recently Wands {\em et al.} \cite{Wands:2002bn} rederived the 
first of the consistency relations (the multi-field version of the standard 
single field consistency realtion) assuming slow-roll only {\em at} horizon 
crossing.  On the other hand the slow-roll approximation during whole stage 
of inflation is required to obtain the second consistency relation [we 
will explain this issue in the next section].

In this work we shall consider the more general situation where the 
slow-roll conditions are not necessarily satisfied even at horizon crossing 
and check the validity of the two 
consistency relations numerically in ``realistic" double inflation 
models.  
The models we adopt are the double inflation with two massive scalar 
fields (both noninteracting 
\cite{Polarski:1992dq,Polarski:1994rz,Langlois:dw} and interacting 
\cite{Linde:1996gt}) and the two-stage supersymmetric inflation with 
tachyonic (spinodal) instability \cite{Linde:1993cn,Copeland:1994vg,RSG} 
where the second derivative of the potential becomes negative.

The former model is probably the simplest double-inflation generalisation 
of the chaotic inflationary scenario.  The second model is motivated by 
supersymmetric theories \cite{Dvali:ms}-\cite{Linde:1997sj}, in which case 
the potentials of scalar fields generically have tachyonic instability 
regions.  Since these two kinds of models include the basic properties of 
double inflation, it is straightforward to extend our analysis to other 
double inflationary scenarios.

We organise our paper as follows.
In Sec.~\ref{general} we present the general framework of our analysis 
including
the multi-field decomposition into adiabatic and entropy field 
perturbations 
and the resulting power spectra of correlated density 
perturbations.  We also discuss the limitation of the slow-roll 
approximation
in the multi-field context.
In Sec.~\ref{tmassive} we analyze the model with two massive scalar 
fields.  
Sec.~\ref{supergra} is devoted to the double inflation with a tachyonic 
instability while the final section concludes.

\section{General Framework} \label{general}

Let us consider two-field inflation with minimally coupled scalar
fields, $\phi$ and $\chi$, with a potential $V(\phi, \chi)$.  
In a flat Friedmann-Lemaitre-Robertson-Walker (FLRW) background 
with a scale factor $a$, the background equations are 
\begin{eqnarray}
\label{back2}
& & H^2 \equiv \left(\frac{\dot{a}}{a}\right)^2=\frac{\kappa^2}{3} 
\left(\frac12 \dot{\phi}^2+ \frac12 \dot{\chi}^2+V \right)\,,~~~ 
\dot{H}=-\frac{\kappa^2}{2} \left( \dot{\phi}^2 +\dot{\chi}^2 \right)\,, \\
& &\ddot{\phi}+3H\dot{\phi}+V_{\phi}=0\,,~~~
\ddot{\chi}+3H\dot{\chi}+V_{\chi}=0\,,
\label{back}
\end{eqnarray}
where $V_{\phi} \equiv \partial V/\partial \phi$, 
$H$ is the Hubble expansion rate, and 
$\kappa^2=8\pi/M_p^2$ with $M_p$ being the Planck mass.  At linear 
order minimally coupled scalar fields do not induce an anisotropic stress 
\cite{KS,MFB,NPB} and hence scalar metric perturbations can be 
characterised by a single potential $\Phi$.  The metric in longitudinal 
gauge then becomes: 
\beqn
ds^2=-(1+2\Phi)dt^2+a^2(1-2\Phi)\delta_{ij} dx^i dx^j\,\,.
\label{metric}
\eeqn
The Fourier transformed, 
linearised Einstein
equations for field and metric perturbations in this gauge are 
\begin{eqnarray} \label{Phi1} 
& & \dot{\Phi} + H\Phi = \frac{\kappa^2}{2} 
\left( \dot{\phi}\delta 
\phi+\dot{\chi}\delta \chi \right) \,, \\
& & \delta\ddot{\phi} + 3H\delta\dot{\phi}+
\left( \frac{k^2}{a^2} + V_{\phi\phi} \right) \delta\phi= 
-2V_{\phi}\Phi+ 4\dot{\phi}\dot{\Phi}-V_{\phi\chi}
\delta \chi \,, \\
& & \delta\ddot{\chi} + 3H\delta\dot{\chi}+ \left( \frac{k^2}{a^2} + 
V_{\chi\chi} \right) \delta\chi= -2V_{\chi} \Phi+ 4\dot{\chi}\dot{\Phi}
-V_{\phi\chi} \delta \phi \,,
\label{perturbed}
\end{eqnarray}
where $k$ is comoving momentum (wavenumber). All first order quantities in 
the 
equations that follow are functions of both $k$ and $t$ (the $k$ subscript 
is 
implicit) \footnote{In this paper we will often use the phrase ``horizon 
crossing". This
should be read ``Hubble radius crossing" occurring for a mode with 
wavenumber $k$ when $k = aH$.}.

We now provide a self-contained review of the 
decomposition of adiabatic  and isocurvature scalar field 
perturbations \cite{Gordon:2000hv}  and the resulting 
spectra of correlated perturbations \cite{Bartolo:2001rt}. 
These two papers are our basic references in this section and we will, 
where 
possible, follow their notation.

We will then also
discuss the limitations of results obtained using slow-roll analysis.  

Let us first introduce the ``adiabatic'' field, $\sigma$, and 
the ``entropy'' field, $s$, defined by 
\beqn
d\sigma = (\cos \theta)d\phi +(\sin \theta)d\chi,~~~~
ds = -(\sin \theta) d\phi +(\cos \theta) d\chi\,.
\label{sigs}
\eeqn
Here $\theta$ is the angle of the trajectory in field space,
satisfying $\tan \theta=\dot{\chi}/\dot{\phi}$.  
With an effective potential $V(\phi,\chi)$, the equations 
for adiabatic and entropy field perturbations are written in the form 
\cite{Gordon:2000hv} 
\beqn
\delta \ddot{\sigma}+3H\delta \dot{\sigma}+\left( \frac{k^2}{a^2}
+V_{\sigma \sigma}-\dot{\theta}^2 \right)\delta \sigma &= &
-2V_{\sigma}\Phi+4\dot{\sigma}\dot{\Phi}+ 2(\dot{\theta}\delta 
s)^{\bullet}-\frac{2V_{\sigma}} {\dot{\sigma}} \dot{\theta} \delta s, \\
\delta \ddot{s}+3H\delta \dot{s}+\left( \frac{k^2}{a^2}
+V_{ss}+3\dot{\theta}^2 \right)\delta s 
&=& \frac{\dot{\theta}}{\dot{\sigma}}
\frac{k^2}{2\pi G a^2}\Phi\,,
\label{deltas}
\eeqn
where 
\beqn
V_{\sigma \sigma} &=& (\cos^2 \theta) V_{\phi \phi}
+(\sin 2\theta)V_{\phi \chi}+(\sin^2 \theta) V_{\chi \chi},\\
V_{ss} &=& (\sin^2 \theta) V_{\phi \phi}
-(\sin 2\theta)V_{\phi \chi}+(\cos^2 \theta) V_{\chi \chi}\,.
\label{Vdd}
\eeqn
{}From eq.~(\ref{Phi1}) we have
\beqn
\Phi=\frac{\kappa^2}{2a} \int a\dot{\sigma} \delta \sigma dt\,.
\label{Phisource}
\eeqn
This indicates that the gravitational potential is sourced by the 
adiabatic field perturbation.

Introducing the Sasaki-Mukhanov variable \cite{SaMuka} 
\beqn
Q_\sigma \equiv \delta\sigma+\frac{\dot{\sigma}}{H}\Phi,
\label{SM}
\eeqn
the equation for the adiabatic field perturbation 
can be rewritten as \cite{Gordon:2000hv} 
\beqn
\ddot{Q}_{\sigma}+3H\dot{Q}_{\sigma}+ \left[
 \frac{k^2}{a^2} +V_{\sigma 
\sigma}-\dot{\theta}^2 
-\frac{\kappa^2}{a^3}\left(\frac{a^3\dot{\sigma}^2}
{H}\right)^{\bullet} \right] 
Q_{\sigma}= 2(\dot{\theta}\delta s)^{\bullet}-2\left(\frac{V_{\sigma}} 
{\dot{\sigma}} +\frac{\dot{H}}{H}\right)\dot{\theta} \delta s\,.
\label{SMeq}
\eeqn
The slow-roll solutions for $Q_{\sigma}$ and $\delta s$ can be
obtained by neglecting the second-order derivatives ($\ddot{Q}_\sigma$ and 
$\delta\ddot{s}$) in eqs.~(\ref{SMeq}) and (\ref{deltas}).  The
evolution of fluctuations using this slow-roll approximation shows
fairly good agreement with numerical results except around the end 
of inflation \cite{Starobinsky:2001xq} unless there exists an
intermediate non-inflationary stage (see ref.~\cite{Polarski:1992dq}).  
Other kinds of slow-roll approximations discussed
later are more problematic however.  

Note, however, that neglecting the
second-order derivatives in eqs.~(\ref{SMeq}) and (\ref{deltas}) still
leads to deviation of the power spectra at the {\it end} of
inflation as found in numerical simulations in
ref.~\cite{Starobinsky:2001xq}.  In this work, we numerically follow
the evolution of perturbations during double inflation and estimate
the spectra right after the end of inflation.  

To provide the comparison to our full numerical results consider the 
solutions for eqs.~(\ref{SMeq}) and (\ref{deltas}), found by  neglecting 
$\ddot{Q}_\sigma$ and $\delta\ddot{s}$ \cite{Bartolo:2001rt}.  
These solutions correspond to neglecting the decaying modes 
of $Q_{\sigma}$ and $\delta s$. Then one has 
\beqn
Q_{\sigma}\simeq Af(t)+BP(t)\,,~~~~\delta s \simeq Bg(t)\,.
\label{slowsolu}
\eeqn
Here $A = A(k)$ and $B = B(k)$.
When $f=g=1$ and $P=0$ at horizon crossing ($k=aH$) the amplitudes $A$ and 
$B$ 
are determined by the quantum fluctuations within the Hubble radius: 
\beqn
A=\frac{H_k}{\sqrt{2k^3}}e_{Q}({\bf k})\,,~~~ 
B=\frac{H_k}{\sqrt{2k^3}}e_{s}({\bf k})\,.
\label{AB}
\eeqn
Here $e_{Q}({\bf k})$ and $e_{s}({\bf k})$ are classical stochastic 
Gaussian quantities, satisfying $\langle e_{Q} ({\bf k}) \rangle = \langle 
e_{s} ({\bf k}) \rangle=0$ and $\langle e_i ({\bf k}) e^*_{j}({\bf k'}) 
\rangle=\delta_{ij} \delta^{(3)}({\bf k}-{\bf k'})$.  Note that $H_k$ is 
the Hubble parameter at horizon crossing.
We caution the reader that in the context of double inflation $P$ can 
be nonzero at horizon crossing due to strong correlations. Clearly then 
the 
assumption of uncorrelated adiabatic and entropy perturbations at $k=aH$
are not generally justified.  In order to make an accurate 
numerical analysis we choose the Bunch-Davies vacuum state deep inside the 
horizon ($k \gg aH$) so that the $\dot{\theta}$ term in the rhs of 
eq.~(\ref{SMeq}) is negligible initially.

On super-Hubble scales ($k \ll aH$) the slow-roll solution for $\delta s$
can be written as 
\beqn
g(t)=\exp \left( -\int_{N(t)}^{N_k}
\frac{\mu_s^2}{3H^2}\,dN \right) \simeq
\exp \left[ -\frac{\mu_s^2}{3H^2} 
\left(N_k-N(t)\right) \right]\,,
\label{deltas_slow}
\eeqn
where $\mu_s^2 \equiv V_{ss}+3\dot{\theta}^2$ and 
$N(t)=-\int_{t_f}^t H dt$ with $t_f$ being the time at the end of 
inflation. The quantity, $N_k=-\int_{t_f}^{t_k} H dt$, corresponds to the 
e-folding between the horizon crossing and the end of inflation.

In deriving eq.~(\ref{deltas_slow}) the time-dependence of the 
$-\mu_s^2/(3H^2)$ term has been neglected, and this term is 
pulled out of the integral.  In the single-field inflationary scenario, 
the variation of this term is associated with the end of inflation, in 
which 
case the error in this approximation is not significant for cosmologically 
relevant scales.  In the case of 
double inflation, the situation is quite different.  Since the mass 
term $-\mu_s^2/(3H^2)$ already grows large at the end of the {\it first} 
stage of inflation, the assumption that the value of $-\mu_s^2/(3H^2)$ 
will 
not change during {\it both} stages of inflation is not 
generally valid.  In fact we shall numerically show later that this term 
typically changes significantly during double-inflation. This  
casts doubts on results derived using this approximation and suggests that 
a 
more sophisticated approximation may be needed to handle multiple phases 
of 
inflation completely.

The slow-roll expansion for $-\mu_s^2/(3H^2)$ is 
given as \cite{Bartolo:2001rt} 
\beqn
-\frac{\mu_s^2}{3H^2} =
-\frac{\epsilon_{\chi}\eta_{\phi\phi}+ 
\epsilon_{\phi}\eta_{\chi\chi}}{\epsilon_t} +2\frac{(\pm 
\sqrt{\epsilon_\phi})(\pm \sqrt{\epsilon_\chi})} 
{\epsilon_t}\eta_{\phi\chi},\,
\label{mu}
\eeqn
where the slow-roll parameters are defined by
\beqn
\epsilon_I \equiv \frac{1}{2\kappa^2} \left( \frac{V_{\phi_I}}{V} 
\right)^2\,,~~~~~ \eta_{IJ} \equiv 
\frac{1}{\kappa^2} \frac{V_{\phi_I \phi_J}}{V}\,,
\label{slowpara}
\eeqn
with $\epsilon_t \equiv \epsilon_{\phi}+\epsilon_{\chi}$.
The entropy field perturbation at the end of inflation is approximately 
expressed as eq.~(\ref{deltas_slow}) with 
\beqn
g(t_f)=\exp\left[\left(-\frac{\epsilon_{\chi}\eta_{\phi\phi}+ 
\epsilon_{\phi}\eta_{\chi\chi}}{\epsilon_t} +2\frac{(\pm 
\sqrt{\epsilon_\phi})(\pm \sqrt{\epsilon_\chi})} 
{\epsilon_t}\eta_{\phi\chi}\right)_k N_k\right]\,,
\label{g}
\eeqn
where we set $N(t_f)=0$.  The slow-roll parameters in this
expression is evaluated at horizon crossing, $k=aH$, 
since the constancy of the mass term is assumed 
in eq.~(\ref{deltas_slow}) [the subscript ``$k$" in eq.~(\ref{g}) 
denotes the value at horizon crossing].

The slow-roll solution for $Q_{\sigma}$ at the end of inflation
can be obtained by assuming the constancy of $\mu_Q^2/H^2 \equiv 
\left(V_{\sigma \sigma}-\dot{\theta}^2 
-\kappa^2 a^{-3} (a^3\dot{\sigma}^2/H)^{\bullet}\right)/H^2$ and 
$\dot{\theta}/H$, as 
\beqn
f(t_f)=\exp\left[\left(-\frac{\epsilon_{\chi}\eta_{\chi\chi}+ 
\epsilon_{\phi}\eta_{\phi\phi}}{\epsilon_t}-2\frac{(\pm 
\sqrt{\epsilon_\phi})(\pm \sqrt{\epsilon_\chi})} 
{\epsilon_t} \eta_{\phi\chi}+2\epsilon_t \right)_k N_k\right]\,,~~~~
P(t_f)=2\,g(t_f) \left(\frac{\dot{\theta}}{H}\right)_k \frac{e^{\zeta_k 
N_k}-1}{\zeta_k} \,,
\label{Qslow_roll}
\eeqn
where
\beqn
\zeta \equiv \frac{\mu_s^2-\mu_Q^2}{3H^2} = 
\frac{(\epsilon_{\phi}-\epsilon_{\chi})(\eta_{\chi\chi}-\eta_{\phi\phi})} 
{\epsilon_t}-4\frac{(\pm \sqrt{\epsilon_\phi})(\pm \sqrt{\epsilon_\chi})} 
{\epsilon_t}\eta_{\phi\chi} +2\epsilon_t \,,
\label{D_def}
\eeqn
and  
\beqn
\frac{\dot{\theta}}{H}=\frac{\epsilon_{\phi}-\epsilon_{\chi}}
{\epsilon_t}\eta_{\phi\chi}+\frac{(\pm \sqrt{\epsilon_\phi})(\pm \sqrt
{\epsilon_\chi})} {\epsilon_t} (\eta_{\phi\phi}-\eta_{\chi\chi}) \,.
\label{dtheta}
\eeqn 
In eq.~(\ref{Qslow_roll}) $\zeta_k$ and $(\dot{\theta}/H)_k$ are 
evaluated at horizon crossing due to the assumption of time-independence
during inflation.  This assumption is not generally justified in the 
context of the double inflation, as we already mentioned.

The curvature perturbation, ${\cal R}$, is defined 
by \cite{Gordon:2000hv} 
\beqn
{\cal R} \equiv \Phi+H \frac{\dot{\phi}\delta\phi+\dot{\chi}\delta\chi} 
{\dot{\phi}^2+\dot{\chi}^2}=\frac{H}{\dot{\sigma}} Q_{\sigma} \,.
\label{cal R}
\eeqn
Since the time-derivative of ${\cal R}$ is given 
as \cite{SaMuka,Gordon:2000hv} 
\beqn
\dot{\cal R}=\frac{H}{\dot{H}}\frac{k^2}{a^2}\Phi+
\frac{2H}{\dot{\sigma}} \dot{\theta}\delta s \,,
\label{dotR}
\eeqn
the curvature perturbation is not conserved even in the large-scale
limit ($k \to 0$) in the presence of the entropy field perturbation,
$\delta s$.  Therefore the constancy of ${\cal R}$ that is typically
assumed in the slow-roll single-field inflationary scenario is not valid 
in 
the multi-field case.  Instead we need to estimate the power spectrum of 
${\cal R}$ at the end of inflation from eq.~(\ref{slowsolu}), as 
\beqn
{\cal P}_{\cal R}=\left( \frac{H_k}{2\pi} \right)^2 
\frac{H^2(t_f)}{\dot{\sigma}^2(t_f)} 
\left[ |f^2(t_f)|+|P^2(t_f)| \right]\simeq 
\frac{1}{\pi} \left( \frac{H_k}{M_p} \right)^2 
\frac{1}{\epsilon_t(t_f)} \left[ |f^2(t_f)|+|P^2(t_f)| \right] \,.
\label{P_R}
\eeqn

The isocurvature perturbation of two scalar fields $\chi$ and $\phi$ is 
defined 
by \cite{KS} 
\beqn
S_{\chi \phi} \equiv \frac{\delta \rho_{\chi}}{\rho_{\chi}+p_{\chi}}- 
\frac{\delta \rho_{\phi}}{\rho_{\phi}+p_{\phi}} 
=\dot{\delta}_{\chi \phi}-3H\delta_{\chi \phi} \,,
\label{iso}
\eeqn
where $\delta_{\chi \phi} \equiv \delta \chi/\dot{\chi}-\delta 
\phi/\dot{\phi}=\dot{\sigma}/(\dot{\phi}\dot{\chi}) \delta s$.  
Neglecting the contribution from 
the $\delta \dot{s}$ term, the isocurvature perturbation can be written in 
terms of the entropy field perturbation, $\delta s$, as 
\beqn
S_{\chi \phi} = T_{\chi\phi}\delta s\,,~~~~{\rm with}~~~~
T_{\chi\phi} \simeq 
-3\frac{\sqrt{4\pi}}{M_p} \frac{\sqrt{\epsilon_t}} {(\pm 
\sqrt{\epsilon_\phi}) (\pm \sqrt{\epsilon_\chi})} \,.
\label{S2}
\eeqn
We note that when the slow-roll conditions are violated, the 
$\delta \dot{s}$ term
may provide a contribution to the isocurvature perturbation that is not 
captured by eq.~(\ref{S2}) which can induce small differences when 
compared with the definition (\ref{iso}).

Making use of eq.~(\ref{S2}), the power spectrum of the isocurvature 
perturbation at the end of inflation is found to be 
\beqn
{\cal P}_{S}=
\left( \frac{H_k}{2\pi} \right)^2 T_{\chi\phi}^2 |g^2(t_f)|
\simeq \frac{9}{\pi} \left( \frac{H_k}{M_p} \right)^2 
\frac{\epsilon_t(t_f)}{\epsilon_{\phi}(t_f) \epsilon_{\chi}(t_f) } 
|g^2(t_f)| \,.
\label{P_S}
\eeqn
The cross-spectrum between $Q_{\sigma}$ and $\delta s$ is estimated 
as $P_{Q\delta s}=(H_k/2\pi)^2 g(t) P(t)$ from eq.~(\ref{slowsolu}).  
Then we find the cross-spectrum between ${\cal R}$ and $S$ as
\beqn
{\cal P}_C=
\left( \frac{H_k}{2\pi} \right)^2 \frac{H(t_f)}{\dot{\sigma}(t_f)}
\,T_{\chi\phi}
\,g(t_f) P(t_f) \simeq -\frac{6}{\pi} 
\left(\frac{\dot{\theta}}{H}\right)_k 
\left(\frac{H_k}{M_p}\right)^2 \frac{e^{\zeta_k N_k}-1}{\zeta_k} 
\frac{|g^2(t_f)|}{(\pm \sqrt{\epsilon_\phi(t_f)})(\pm 
\sqrt{\epsilon_\chi(t_f)})} \,.
\label{P_C}
\eeqn

The spectral indices for the power-spectrum, ${\cal P}$, is defined by 
\beqn
n-1 \equiv \frac{d\,{\rm ln} {\cal P}}{d\,{\rm ln} k}=
(1+\epsilon_t) \frac{d\,{\rm ln} {\cal P}}{d\,{\rm ln} a}
\Biggr|_{k=aH}  \,.
\label{n}
\eeqn
Therefore the spectral indices for ${\cal P}_{\cal R}$, ${\cal P}_S$, and 
${\cal P}_C$ read \cite{Bartolo:2001rt} 
\beqn \label{tilt} 
n_{\cal R}-1 &=& -6\epsilon_t+2\frac{\epsilon_\phi 
\eta_{\phi\phi}+\epsilon_\chi \eta_{\chi\chi}} {\epsilon_t} +4\frac{(\pm 
\sqrt{\epsilon_\phi}) (\pm \sqrt{\epsilon_\chi})}{\epsilon_t} 
\eta_{\phi\chi} 
-\frac{8|f^2(t_f)|}{|f^2(t_f)|+|P^2(t_f)|} 
\left(\frac{\dot{\theta}}{H}\right)_k^2 \frac{e^{ -\zeta_k 
N_k}}{\zeta_k}(1-e^{-\zeta_k N_k}) \,, \\
\label{tiltS2} 
n_S-1 &=& -2\epsilon_t+2\frac{\epsilon_\phi 
\eta_{\chi\chi}+\epsilon_\chi \eta_{\phi\phi}} {\epsilon_t} -4\frac{(\pm 
\sqrt{\epsilon_\phi}) (\pm \sqrt{\epsilon_\chi})}{\epsilon_t} 
\eta_{\phi\chi} \,, \\
\label{tiltS} 
n_C-1 &=& -2\epsilon_t+2\frac{\epsilon_\phi 
\eta_{\chi\chi}+\epsilon_\chi \eta_{\phi\phi}} {\epsilon_t} -4\frac{(\pm 
\sqrt{\epsilon_\phi}) (\pm \sqrt{\epsilon_\chi})}{\epsilon_t} 
\eta_{\phi\chi}
-\frac{\zeta_k e^{\zeta_k N_k}}{e^{\zeta_k N_k}-1} \,, 
\eeqn 
where the slow-roll parameters are evaluated at horizon crossing. 
The spectrum $P_T$ and the spectral index $n_T$ of tensor 
perturbations are calculated by analyzing the equation of 
massless gravitational fields \cite{Lyth:1998xn}: 
\beqn \label{GW} 
P_T=\left( \frac{4}{\sqrt{\pi}} \frac{H_k}{M_p} \right)^2 
\,,~~~~n_T=-\frac{8\pi}{M_p^2} \left(\frac{\dot{\sigma}}{H}
\right)_k^2 \,.  
\eeqn 
We introduce two ratios $r_C$ and $r_T$, which are defined as 
\beqn \label{r_C} 
r_C \equiv \frac{P_C}{\sqrt{P_{\cal R}P_S}}\,, 
\eeqn 
and
\beqn \label{r_T} 
r_T \equiv \frac{P_T}{16P_{\cal R}}\,.
\eeqn 
{}From eqs.~(\ref{P_R}), (\ref{P_S}) and (\ref{P_C}) we find
that the correlation ratio $r_C$ can be expressed as 
\beqn \label{x} 
r_C=\frac{x}{\sqrt{1+x^2}}\,,~~~~ {\rm with}~~~~
x= \frac{P(t_f)}{f(t_f)} \,.  
\eeqn 
Therefore $r_C^2$ lies in the range $0 \le r_C^2 \le 1$.  Note that the 
relation (\ref{x}) is obtained without assuming that the 
adiabatic/entropy masses and $\dot{\theta}/H$ are constant 
after horizon crossing. Namely, the equality $\simeq$ 
in eqs.~(\ref{P_R}), (\ref{P_S}) and 
(\ref{P_C}) is not used when we derive eq.~(\ref{x}).  If the slow-roll 
solutions (\ref{Qslow_roll}) are employed, we have 
\beqn \label{xslow} 
x \simeq 2\left(\frac{\dot{\theta}}{H}\right)_k 
\frac{1-e^{-\zeta_k N_k}}{\zeta_k} \,. 
\eeqn 
The behaviour of the term $\dot{\theta}/H$ is 
most important when we analyze the correlation between adiabatic and 
isocurvature perturbations.  In eq.~(\ref{xslow}) the ``frozen" value of 
$\dot{\theta}/H$ is used at horizon crossing.  However, since the 
assumption of constant $\dot{\theta}/H$ is not generally valid during 
double inflation, the slow-roll result (\ref{xslow}) leads to some errors 
in estimating $r_C$ at the end of double inflation.
When $\dot{\theta}/H$ varies significantly, we have to integrate this term 
from first horizon crossing to the end of inflation rather than use the 
``frozen" value at horizon crossing.  Note that if $\dot{\theta}/H$ is 
vanishingly small during the {\it both} phases of inflation the 
correlation vanishes ($r_C=0$). 

The tensor to scalar ratio $r_T$ can be evaluated 
without using the slow-roll equality in eqs.~(\ref{P_R}) and (\ref{GW}), as
\beqn \label{r_T2} 
r_T=\frac{4\pi}{M_p^2}\left(\frac{\dot{\sigma}(t_f)}{H(t_f)} 
\right)^2 \frac{1}{|f^2(t_f)|+|P^2(t_f)|} =\frac{4\pi}{M_p^2} 
\left(\frac{\dot{\sigma}}{H}\right)^2_k \frac{1}{1+x^2} \,.  \eeqn %
Here we used the fact that $(H/\dot{\sigma})f$ is conserved after horizon 
crossing, i.e., $(H/\dot{\sigma})_k=(H(t_f)/\dot{\sigma}(t_f))\,f(t_f)$
[see eq.~(\ref{dotR}) with $k \ll aH$ and $\delta s=0$].
Making use of eqs.~(\ref{GW}), (\ref{x}) and (\ref{r_T2})
we get the consistency relation 
\beqn \label{consistency1} 
r_T=-\frac{n_T}{2}\left(1-r_C^2 \right) \,. 
\eeqn 
This indicates that the correlation between adiabatic and isocurvature 
perturbations leads to the modification of the consistency relation in the 
single field case ($r_T=-n_T/2$).

In deriving eq.~(\ref{consistency1}), we did not exploit the assumption 
that the adiabatic/entropy mass and $\dot{\theta}/H$ are constant 
after horizon crossing.  Then this consistency relation should be valid as 
long as the slow-roll conditions are satisfied {\it at horizon crossing}, 
in which case the uncorrelated solutions for $Q_{\sigma}$ and $\delta s$ 
can be used at $k=aH$ \cite{Wands:2002bn} \footnote{Note that
the decaying mode for ${\cal R}$ can be important in some non slow-roll 
inflationary scenarios \cite{Starobinsky:ts,Leach:2001zf}.  In this case 
the second derivatives of eqs.~(\ref{deltas}) and (\ref{SMeq}) are not 
necessarily small and the first term in the rhs of eq.~(\ref{slowsolu}) is 
not negligible.  Then we need to add the decaying mode solutions to 
eq.~(\ref{slowsolu}).  The consistency relation (\ref{consistency1}) does 
not cover this case, although the enhancement of the decaying mode occurs 
only in some restricted situations \cite{Starobinsky:ts,Leach:2001zf}.}.  
In the context of double inflation there are some cases where the 
slow-roll 
conditions can be violated at horizon crossing, implying that the 
consistency relation (\ref{consistency1}) does not hold automatically when 
applied to realistic double inflation models.

The authors in ref.~\cite{Bartolo:2001rt} obtained the following second 
consistency relation from the slow-roll results (\ref{tilt})-(\ref{tiltS})
together with (\ref{GW}) and (\ref{r_T2}), as 
\beqn \label{consistency2} 
\left(n_C-n_S\right)r_T= -\frac{n_T}{4}\left(2n_C-n_{\cal R}-n_S 
\right)\,. 
\eeqn 
Note that the constancy of the adiabatic/entropy mass and $\dot{\theta}/H$ 
is 
assumed in deriving this relation.
Therefore it is likely that the second consistency relation 
(\ref{consistency2}) is more strongly affected by the violation of the 
slow-roll conditions compared to the first consistency relation 
(\ref{consistency1}).

While the slow-roll results which include the quantities $n_{\cal R}$, 
$n_S$, 
and $n_C$ can exhibit strong deviation from the numerical results, 
the spectral index $n_T$ of the gravitational wave is well described by  
eq.~(\ref{GW}) even in the context of double inflation.  Therefore 
provided 
that the correlation is small at horizon crossing, the first consistency 
relation (\ref{consistency1}) is expected to be reliable as long as we use 
$x$ in eq.~(\ref{x}) instead of the slow-roll result in eq.~(\ref{xslow}).

In the following section we shall compare the above formula with full 
numerical simulations for concrete
models of double inflation (see Appendix for the numerical method 
to evaluate power spectra and correlations).  We will provide a detailed 
analysis of the spectra of perturbations and the validity of the 
consistency relations derived from the above analysis.  We will also 
discuss the parameter ranges where the correlation of adiabatic and 
isocurvature perturbations is strong.

\section{Double inflation with two massive scalar fields}
 \label{tmassive}

Let us first consider a simple model where massive scalar fields, $\phi$ 
and $\chi$, are coupled through an interaction term $\frac12 
g^2\phi^2\chi^2$: 
\beqn \label{V}
V(\phi,\chi)=\frac12 m_{\phi}^2\phi^2+\frac12 m_{\chi}^2
\chi^2 + \frac12 
g^2\phi^2\chi^2\,. 
\eeqn 
There are three parameters associated with this potential: 
$m_{\phi}$, $m_{\chi}$ and $g$.  Then there are four free parameters
associated with the initial conditions of the fields: $\phi_{i}$,
$\chi_{i}$, $\dot\phi_{i}$ and $\dot\chi_{i}$.  Making use of the slow
roll approximation, $\dot\phi = -V_{\phi}/3H$ and $\dot\chi =
-V_{\chi}/3H$ with $H^2=(8\pi/3M_p^2)V$ in eqs.~(\ref{back2}) and
(\ref{back}), the initial conditions of $\dot\phi$ and $\dot\chi$ are
determined by $\phi_i$ and $\chi_i$
\footnote{Clearly assuming slow-roll to set the initial conditions is not 
generally valid. Not assuming this will lead to extra transient violations
of the slow-roll conditions but if inflation is successfully initiated, 
the 
fields should settle to their slow-roll values quickly. At any rate our 
interest is in correlations and violations of the slow-roll approximation 
in a minimal sense. Inverting CMB and LSS data to give information about 
the potential and initial conditions will have to deal with this 
possibility in general however.}.
This assumption cuts down the number of free
parameters to two, $\phi_{i}$ and $\chi_{i}$.  Therefore we have five
free parameters ($m_{\phi}$, $m_{\chi}$, $g$, $\phi_i$ and $\chi_i$)
for the model (\ref{V}).  Once these parameters are given, the
evolution of the background is determined, with the number of e-folds,
$N =-{ \rm ln} (a/a_f)$, with $a_f$ being the value of the scale
factor at the end of inflation \cite{Polarski:1992dq}.  We shall
introduce the number of e-folds, $N_H$, which corresponds to the 
value of $N$ when the scale corresponding to our Hubble radius today 
crossed out the Hubble radius during inflation.  Hereafter we set it to be 
\beqn \label{NH} 
N_H=60 \,, 
\eeqn 
in order to make definite calculations.

\subsection{Non-interacting fields: $g=0$}

In the case where the fields are non-interacting ($g=0$), the slow-roll
approximation in eqs.~(\ref{back2}) and (\ref{back}) gives the relation 
$\phi^2+\chi^2=4N/\kappa^2$. The fields lie on a circle of radius 
$2\sqrt{N}/\kappa$.  Therefore it is useful to write $\phi$ and $\chi$ 
in parametric form \cite{Polarski:1992dq}: 
\beqn
\phi=\frac{2\sqrt{N}}{\kappa}\,\cos \alpha\,,
~~~~ \chi=\frac{2\sqrt{N}}{\kappa}\,\sin \alpha \,.
\label{para}
\eeqn
This means that the evolution of two scalar fields is characterised 
by $N$ and the scalar field position angle, $\alpha$, satisfying
the relation $\tan \alpha=\chi/\phi$.  The field velocity angle, $\theta$, 
defined by eq.~(\ref{sigs}) is related to $\alpha$ by 
\beqn
\tan \theta \simeq -\frac{2m_{\chi}^2\sqrt{N}}
{3H\kappa \dot{\sigma}} \tan \alpha \,.
\label{theal}
\eeqn
Making use of the relation (\ref{para}), we find that 
the number of e-folds can be expressed as \cite{Polarski:1992dq}
\beqn
N = N_0 \frac{(\sin \alpha)^{2/(R^2-1)}}
{(\cos \alpha)^{2R^2/(R^2-1)}} \,,
\label{Napprox}
\eeqn
where 
\beqn
R \equiv m_{\chi}/m_{\phi}\,.
\label{ratio}
\eeqn
Note that the integration constant, $N_0$, roughly 
corresponds to the number of e-folds during the second stage of inflation 
driven by the light scalar field.  Hereafter we shall concentrate on the 
case where the field $\chi$ is heavier than $\phi$, i.e., $R>1$.

In order to know the evolution of the background we need to determine
four parameters: $m_{\phi}$, $R$, $N_0$, and $\alpha$.  
When the total number of e-folds is fixed at around $N_H$, the model 
parameters are reduced to three ($m_{\phi}$, $R$ and $N_0$).  Whether 
inflation is dominated by the heavy or light fields when the scale of 
cosmological relevance crosses the Hubble radius depends on the value of 
$N_0$ relative to $N_H=60$.

Adiabatic perturbations for modes larger than the Hubble radius
during the radiation dominant era can be matched with the curvature 
perturbation at the end of inflation, which are given 
by \cite{Langlois:dw,Gordon:2000hv} 
\begin{eqnarray}
{\cal R} \simeq -\frac{\kappa^2 H(t_*)}{2\sqrt{2k^3}} \left[\phi(t_*) 
e_{\phi}({\bf k})+\chi(t_*) e_{\chi}({\bf k}) \right] =-\frac{\kappa 
H(t_*)\sqrt{N}}{\sqrt{2k^3}} \left[\cos \alpha_* \,e_{\phi}({\bf k})+ \sin 
\alpha_* \,e_{\chi}({\bf k}) \right] \,,
\label{Phirad}
\end{eqnarray}
where $\alpha_*$ is the value of $\alpha$ at the horizon crossing. 
Assuming that the field $\phi$ decays into ordinary matter (baryons, 
photons, neutrinos) and $\chi$ into cold dark matter, super-Hubble 
isocurvature perturbations during the radiation dominant era is expressed 
as \cite{Langlois:dw,Gordon:2000hv} 
\begin{eqnarray}
S \simeq \frac{H(t_*)}{\sqrt{2k^3}} \left[R^2 \frac{e_{\phi}({\bf 
k})}{\phi(t_*)}- \frac{e_{\chi}({\bf k})}{\chi(t_*)} \right] =\frac{\kappa 
H(t_*)}{2\sqrt{N} \sqrt{2k^3}} \left[R^2 \frac{e_{\phi}({\bf k})}{\cos 
\alpha_*}- \frac{e_{\chi}({\bf k})}{\sin \alpha_*} \right] \,.
\label{isorad}
\end{eqnarray}

The expression (\ref{Phirad}) indicates that for the adiabatic 
perturbation the heavy field $\chi$ dominates for $\tan \alpha_*>1$, 
while the light field $\phi$ dominates for $\tan \alpha_*<1$.  {}From 
eq.~(\ref{isorad}) we find that for the isocurvature perturbation the 
heavy 
field $\chi$ dominates for $\tan \alpha_*<1/R^2$, 
while the light field $\phi$ dominates for $\tan \alpha_*>1/R^2$.

Let us estimate the correlation $r_C$ that is derived from the slow-roll 
analysis, see eq.~(\ref{xslow}). This is not actually 
completely valid as we pointed out in the previous section, but useful to 
make rough estimation for the correlation.  We will check, of course, the 
validity of the analytic estimates by numerical simulations.  By a simple 
calculation we find that $x$ defined in eq.~(\ref{xslow}) is given by 
\begin{eqnarray}
x=\frac{R^2(R^2-1)\tan \alpha_* (1+\tan^2\alpha_*)}
{(1+R^2 \tan^2\alpha_*)(1+R^4 \tan^2 \alpha_*)}
\frac{1-e^{-\zeta_k N_k}}{\zeta_k 
N_k}\,.                                                                                                                                                            
\label{x_g=0}
\end{eqnarray}
If the condition, $|\zeta_k| N_k \ll 1$, is satisfied, this reduces to 
\begin{eqnarray}
x=\frac{R^2(R^2-1)\tan \alpha_* (1+\tan^2\alpha_*)}
{(1+R^2 \tan^2\alpha_*)(1+R^4 \tan^2 \alpha_*)}\,.
\label{x_gd}
\end{eqnarray}
Note that when $|\zeta_k| N_k~\gsim~1$ one has 
$|(1-e^{-\zeta_k N_k})/(\zeta_k N_k)|~\simeq 1/|(\zeta_k| N_k)~\lsim~1$.  
Therefore the value of $x$ is smaller than in the case of (\ref{x_gd}).  
Eq.~(\ref{x}) implies that the correlation $r_C$ vanishes for $x=0$ and 
gets larger for increasing $x$.  In particular when $x$ is larger than of 
order unity, the correlation is strong ($r_C$ is close to unity).  {}From 
eq.~(\ref{x_g=0}) we find that there is no correlation if the masses of 
the scalar fields are equal ($R=1$).  We can also make consistency check 
by 
using eq.~(\ref{x_g=0}) or (\ref{x_gd}).  When the masses of the scalar 
fields differ significantly ($R \to 0$ or $R \to \infty$), the correlation 
is also vanishingly small for fixed $\tan \alpha_* $.

In order to discuss the correlation precisely, it is useful to
classify model parameters into three cases \cite{Langlois:dw}: 
(a)~ $\tan \alpha_* \gg 1$, (b)~$\tan \alpha_* 
\ll 1/R^2$ and (c)~$1/R^2<\tan \alpha_* <1$.  
Hereafter we shall analyze the strength of the correlation as well as the 
power spectra and consistency relations, and check the validity of the 
slow-roll analysis.

\vspace{0.3cm}

\begin{center}
 {\bf (a)~$\tan \alpha_* \gg 1$}
\end{center}

In this case the field $\chi$ is the main source for adiabatic 
perturbations, while isocurvature perturbations are dominated by the
field $\phi$.  Therefore both perturbations are regarded as almost
independent ones, and the correlation is weak (see Fig.~\ref{rc}).  In
fact when $\tan \alpha_* \gg 1$ eq.~(\ref{x_gd}) yields 
\begin{eqnarray}
x \simeq \frac{R^2-1}{R^4} \frac{1}{\tan \alpha_*} \,.
\label{x_gd1}
\end{eqnarray}
Therefore the correlation $r_C$ decreases with increasing 
$\tan \alpha_*$ and one has $r_C \to 0$ for $\tan \alpha_* 
\to \infty$.  This decreasing rate is more significant for larger $R$
as can be seen from eq.~(\ref{x_gd1}) and Fig.~\ref{rcnogtan}.

The amplitude of isocurvature perturbations is not typically larger than 
that of adiabatic perturbations unless $\alpha_*$ is so much close to 
$\pi/2$, as shown in Fig.~\ref{spectra}.\footnote{Note, however, that the 
amplitude of isocurvature perturbations can be high if $\alpha_*$ is very 
close to $\pi/2$.}  Since the correlation term in eq.~(\ref{tilt}) is 
neglected and $\epsilon_\phi \ll \epsilon_\chi $ for $\tan \alpha_* \gg 
1$, one has a spectral index of the curvature perturbation that is 
approximately the same as the single field case: 
\begin{eqnarray}
n_{\cal R}-1 \simeq -6\epsilon_\chi+2\eta_{\chi\chi}
=-\frac{1}{\pi} \left( \frac{M_p}{\chi} \right)^2 \,.
\label{nRm}
\end{eqnarray}
 This is a slowly red-tilted spectrum as found in Fig.~\ref{spectra}.
In Fig.~\ref{consist} we plot the ratio $r_T$ 
defined by eq.~(\ref{r_T}) and its value obtained by two consistency 
relations (\ref{consistency1}) and (\ref{consistency2}).  
Except for some discontinuous behaviour which is 
accompanied with numerics\footnote{We evaluated the spectral indices 
numerically using the definition, 
$n=1+\Delta({\rm ln}\,P)/\Delta({\rm ln}\,k)$, which leads to some 
numerical errors and some spikiness in some of the figures.}, 
consistency relations show fairly good agreement 
with the value of the original definition of $r_T$.  In this case since 
$r_C$ is much less than unity the consistency relation (\ref{consistency1}) is 
practically no different from that of the single field case, 
$r_T=-n_T/2$.  
Namely it is almost the same as the single field inflation driven by only 
one scalar field.  Therefore the assumption that $\mu_Q^2/(3H^2)$, 
$\mu_s^2/(3H^2)$ and $\dot{\theta}/H$ do not vary too much during 
inflation can be justified in this case, so not giving strong deviation in the 
consistency relations.

\vspace{0.3cm}

\begin{center}
 {\bf (b)~$\tan \alpha_* \ll 1/R^2$}
\end{center}

In this case the field $\phi$ is the main source for adiabatic
perturbations, while isocurvature perturbations are dominated by the
field $\chi$.  {}From eq.~(\ref{x_gd}) one has 
\begin{eqnarray}
x \simeq R^2(R^2-1) \tan \alpha_* \,,
\label{x_gd2}
\end{eqnarray}
for $R^2 \tan \alpha_* \ll 1$.  Therefore adiabatic and isocurvature 
perturbations are almost independent from each other for smaller $\tan 
\alpha_*$, which can be confirmed in Fig.~\ref{rc}.  In 
Fig.~\ref{rcnogtan} 
we find that the prediction (\ref{x_gd}) overestimates the correlation 
ratio $r_C$ when $\tan \alpha_*$ is small, while eq.~(\ref{x_g=0}) shows 
fairly good agreement with numerical results.  This implies that 
$|\zeta_k| 
N_k$ could be larger than unity, in which case the $(1-e^{-\zeta_k 
N_k})/(\zeta_k N_k)$ term can not be neglected in eq.~(\ref{x_g=0}).

When $\tan \alpha_* \ll 1/R^2$ the amplitude of isocurvature perturbations 
are larger than that of adiabatic ones as predicted by 
eqs.~(\ref{Phirad}) and (\ref{isorad}) [see Fig.~\ref{spectra}].  
The spectrum of curvature perturbations is hardly affected by
isocurvature perturbations because the correlation is small ($r_C \ll 1$).
Therefore the consistency relation in the single-field case should
not be significantly modified in this case.  

In fact, from Fig.~\ref{consist} we find that the first consistency 
relation (\ref{consistency1}) shows good agreement with the original 
definition of $r_T$, while the second one (\ref{consistency2}) 
is not so good.
Indeed we should expect deviations from the predictions of the 
second consistency relation 
around the end of inflation because the masses of the adiabatic/entropy 
fields and $\dot{\theta}/H$ are not constant in this case.  Even in the 
case (a) the discrepancy in the second consistency relation is a bit 
larger 
than in the case of the first one.

\begin{figure}
\begin{center}
\singlefig{7cm}{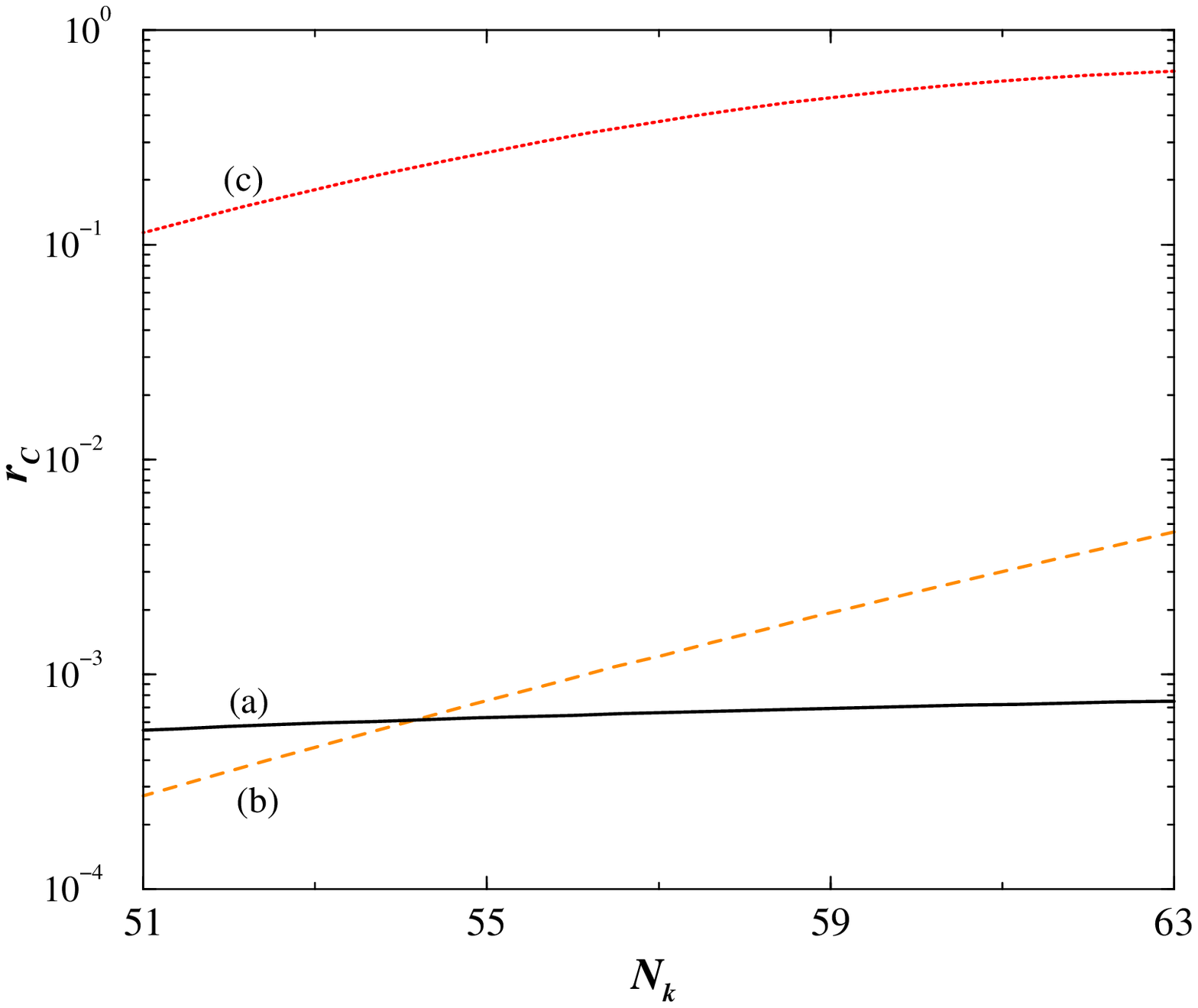}
\begin{figcaption}{rc}{7cm}
Correlation spectra $r_C$ for three different cases
with $R=5$, $m_{\phi}=2.0 \times 10^{-7}M_p$ and $g=0$.  Each case 
corresponds to (a) $\tan \alpha_*=32.0 \gg 1$, (b) $\tan \alpha_*=3.13 \times 
10^{-4} \ll R^{-2}$, and (c) $R^{-2}<\tan \alpha_*=0.16<1$, on the scale 
$N_k=65$.  The case (c) shows strong correlations, while the cases (a) and 
(b) are not.
\end{figcaption}
\end{center}
\end{figure}

\begin{figure}
\begin{center}
\singlefig{10cm}{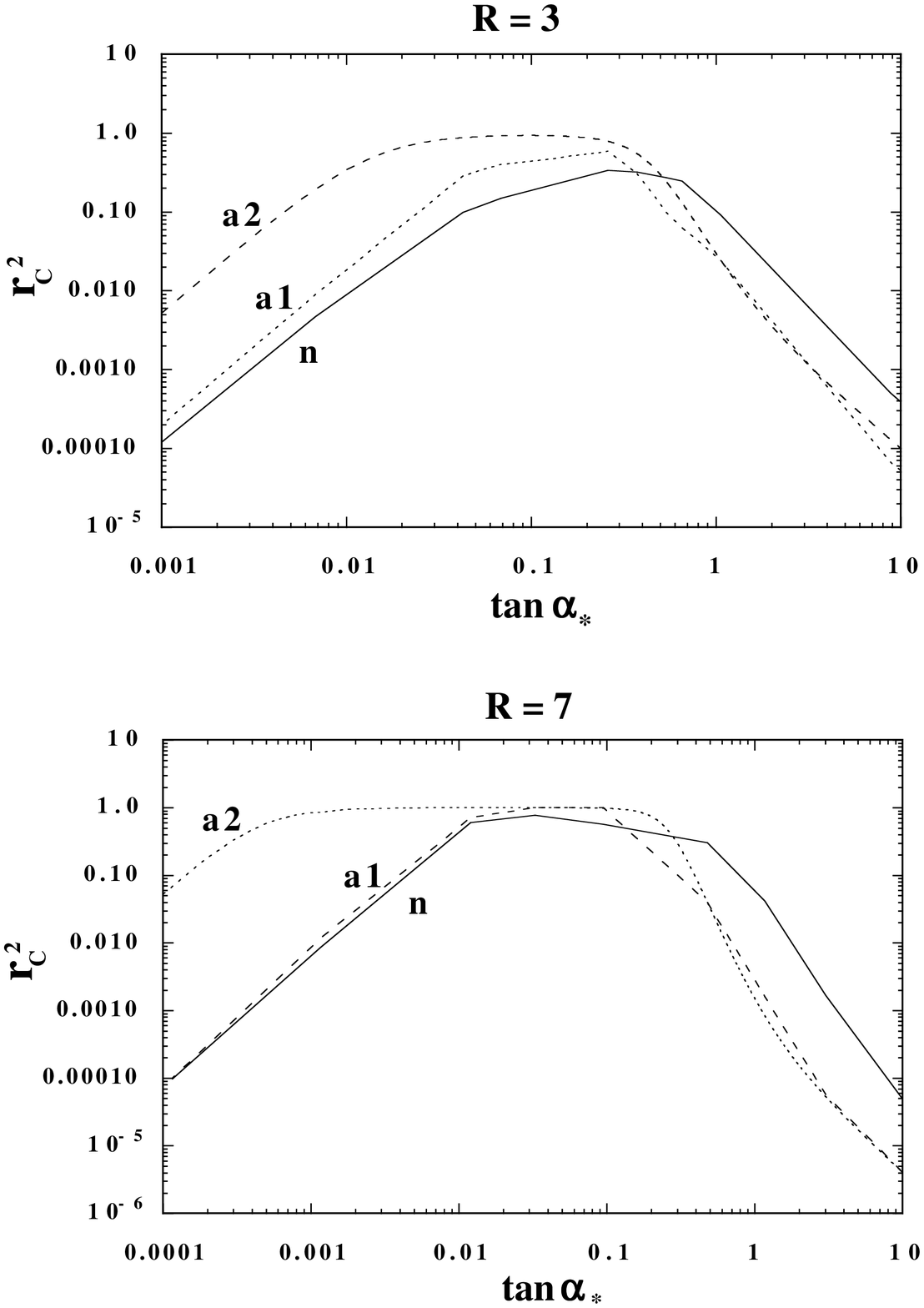}
\begin{figcaption}{rcnogtan}{10cm}
The square of the correlation $r_C$ as a function of $\tan \alpha_*$ 
for $R=3$ and $R=7$ with $m_{\phi}=2.0 \times 10^{-7}M_p$ and $g=0$ 
on the scale corresponding to $N_k=60$.  The solid curve corresponds to the 
numerical result, while dashed (``a1'') and dotted (``a2'') curves 
correspond to the results using eqs.~(\ref{x_g=0}) and (\ref{x_gd}), 
respectively.
\end{figcaption}
\end{center}
\end{figure}

\vspace{0.3cm}

\begin{center}
 {\bf (c)~$1/R^2 \le \tan \alpha_* \le 1$}
\end{center}

In this case both adiabatic and isocurvature perturbations
are sourced by the light field $\phi$, 
but the effect of the heavy field $\chi$ is also important.  
{}From eq.~(\ref{x_gd}) we find 
\begin{eqnarray}
x =\frac{(R^2-1)(R^4+1)}{2R^2(R^2+1)} \,,~~~~
{\rm for}~~~~\tan \alpha_*=\frac{1}{R^2}\,,
\label{x_gd3}
\end{eqnarray}
and 
\begin{eqnarray}
x =\frac{2R^2(R^2-1)}{(R^2+1)(R^4+1)} \,,
~~~~ {\rm for}~~~~\tan \alpha_*=1 \,.
\label{x_gd4}
\end{eqnarray}
Therefore when $\tan \alpha_*=1/R^2$ and $R$ is not too close to
unity, $x$ is typically larger than  unity (for example one has
$x>1.275$ for $R>2$).  In this case the correlation ratio $r_C$ is 
close to 1.  The range of this high correlation gets wider for larger
$R$ as found in Fig.~\ref{rcnogtan}.  When $\tan \alpha_* \simeq 1$,
$x$ is at a maximum  $x_{\rm max} \simeq 0.3$ for $R \simeq
1.7$,  with the correlation ratio ranges $r_C \le 0.28$ in this
case.  As $R$ is increased, the maximum correlation becomes smaller,
as is seen in Fig.~\ref{rcnogtan}.  

Note that we need to include the correction term $(1-e^{-\zeta_k 
N_k})/(\zeta_k N_k)$  in eq.~(\ref{x_g=0}) to accurately estimate the
strength of the correlation.  Fig.~\ref{rcnogtan} clearly indicates
that the correlation is strong around $1/R^2 \le \tan \alpha_* \le 1$.
In this case the correlation term $r_C^2$ is very important in the
consistency relation, (\ref{consistency1}), because $r_C$ will be
close to unity.

As found from Fig.~\ref{rcnogtan} analytic estimates by slow-roll
approximations typically give larger values of $r_C$ around the 
region where the correlation is strong.  When $r_C$ is close to 
unity, this difference can affect the consistency relation
(\ref{consistency1}).
In Figs.~\ref{qmphichi}{phchtheta} we plot the evolution of $\mu_Q^2/(3H^2)$, 
$\mu_s^2/(3H^2)$ and $\dot\theta/H$ for 
$R = 5$, $m_\chi=1 \times 10^{-6} M_p$ and $g=0$ with 
initial conditions $\chi=3M_p$ and $\phi=1.5M_p$.
The heavy field $\chi$ leads to the first phase of inflation until $\tau 
\equiv 10^{-6}M_p t \simeq 20$, which is followed by the 
second stage of inflation driven by $\phi$.
All of $\mu_Q^2/(3H^2)$, $\mu_s^2/(3H^2)$ and $\dot\theta/H$
exhibit rapid increase around the end of the first stage of inflation due 
to the breakdown of the slow-roll conditions for $\chi$.
For example, $\mu_s^2/(3H^2)$ continues to grow by the end 
of the second stage of inflation, whose growth is about $5 \times 10^4$
times compared to its initial value. 

In this case the assumption of the constancy of the mass terms is no
longer justified in eqs.~(\ref{deltas_slow}) and (\ref{Qslow_roll}), 
thereby leading to errors in the correlation $r_C$ if we use the 
estimation in eq.~(\ref{xslow}). In addition, the peak value of 
$\dot{\theta}/H$ typically provides a larger contribution than its value 
at 
horizon crossing in eq.~(\ref{xslow}).  
Therefore we need to evaluate the values of $x$ and $r_C$ numerically
in order to estimate the correlation accurately.

In the case where the correlation is strong at horizon crossing, we
expect to find some deviations even from the predictions of the first 
consistency relation.  In fact the numerical result in Fig.~\ref{consist} 
(c) does not completely agree with the slow-roll results, although 
the deviation is not significant. This case corresponds to the one where the 
slow-roll conditions are violated at horizon crossing.  We have numerically 
checked that the first consistency relation holds well as long as the 
slow-roll conditions are satisfied at horizon crossing, which agrees with 
the claim by Wands {\em et al}~\cite{Wands:2002bn}.  The second consistency 
relation is more strongly affected by the violation of the slow-roll 
conditions during double inflation, especially when the correlation is 
strong.  The slow-roll analysis shows some limitations to correctly 
estimate three spectral indices $n_{\cal R}$, $n_S$, and $n_C$.  Numerical 
analysis is required as well in order to fully understand the strength of 
the correlation and the final power spectra of adiabatic and isocurvature 
perturbations.
 
{}In Fig.~\ref{rc} we find that the correlation is high around 
$N_k~\gsim~60$, and decreases toward smaller scales.  
This corresponds to the 
``light'' inflationary phase with $\theta~\lsim~1/R$ where the 
perturbations are mainly sourced by the field $\phi$ around $N_k \simeq 
60$.  In this case the correlation gets weaker toward smaller scales due 
to the decrease of $\dot{\theta}$.  If the scale $N_k=60$ corresponds 
to the ``heavy'' inflationary phase with $\alpha~\gsim~1/R$, the 
correlation $r_C$ is nearly constant as shown in ref.~\cite{Langlois:dw}.  
This means that $\alpha$ varies slowly during the heavy field inflation, 
which makes $\dot{\theta}$ unsuppressed.  The slow variation of $r_C$ can 
be 
actually found in the case (a) of Fig.~\ref{rc}.  Note that if we choose 
the value of $\alpha$ not much greater than $1/R$ the correlation can be 
higher as claimed in ref.~\cite{Langlois:dw}.

\begin{figure}
\begin{center}
\singlefig{6.5cm}{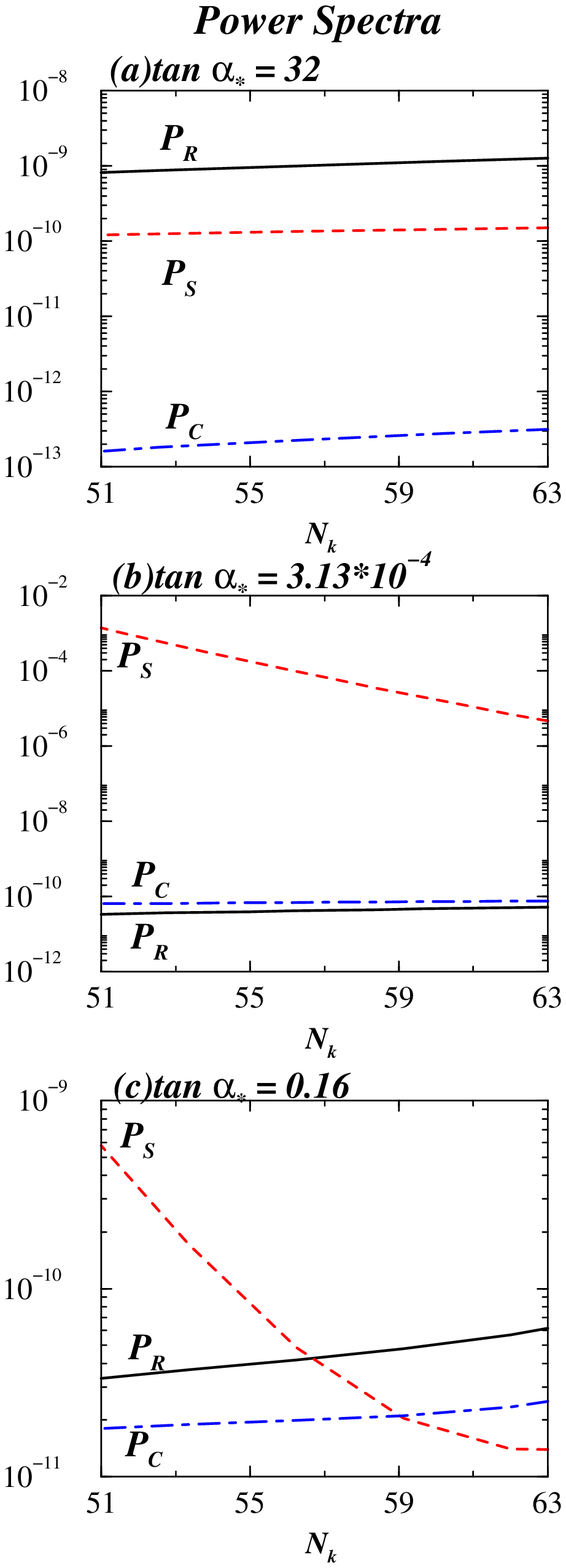}
\begin{figcaption}{spectra}{6.5cm}
The power spectra $P_R$, $P_S$ and $P_C$ 
with $R=5$, $m_{\phi}=2.0 \times 10^{-7}M_p$ and $g=0$.  
The curves correspond to the cases 
(a) $\tan \alpha_*=32.0 \gg 1$ (heavy-field dominated), 
(b) $\tan \alpha_*=3.13 
\times 10^{-4} \ll R^{-2}$ (light-field dominated), and
(c) $R^{-2}<\tan \alpha_*=0.16<1$, on the scale 
corresponding to $N_k=65$ (double inflation).
\end{figcaption}
\end{center}
\end{figure}
\begin{figure}
\begin{center}
\singlefig{6cm}{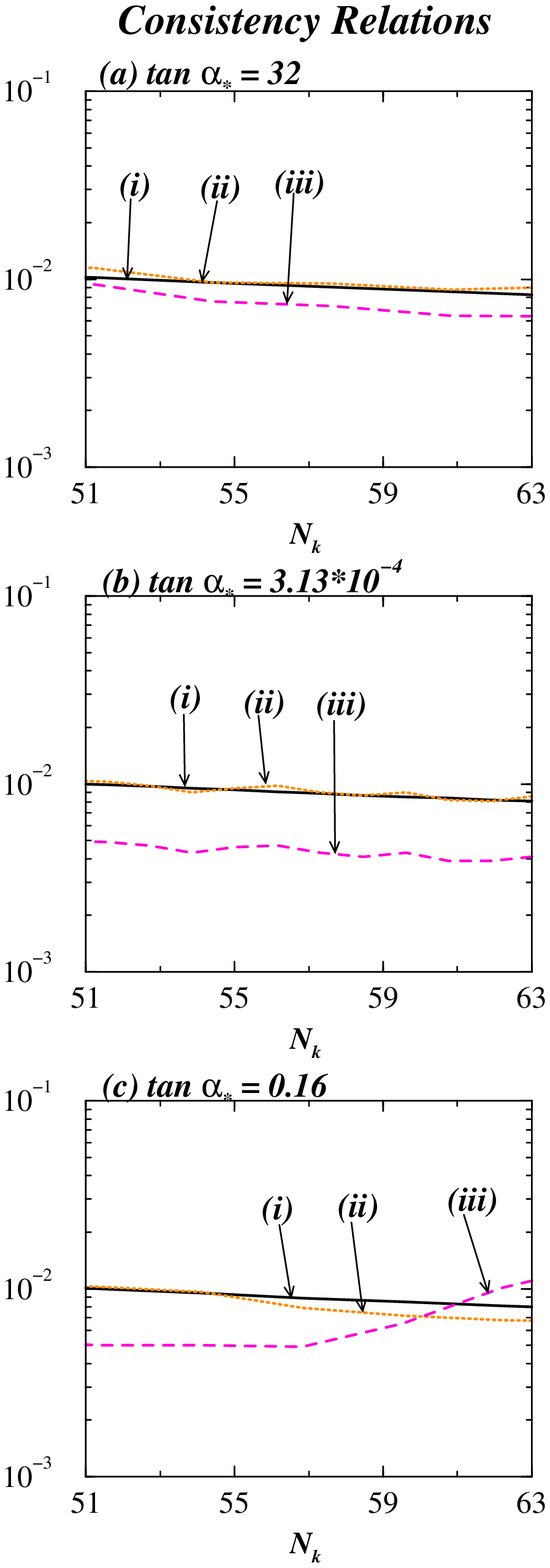}
\begin{figcaption}{consist}{6cm}
The consistency relations
with $R=5$, $m_{\phi}=2.0 \times 10^{-7}M_p$ and $g=0$.  
The curves correspond to the cases 
(a) $\tan \alpha_*=32.0 \gg 1$,
(b) $\tan \alpha_*=3.13 \times 10^{-4} \ll R^{-2}$, and
(c) $R^{-2}<\tan \alpha_*=0.16<1$, on the scale 
corresponding to $N_k=65$ (double inflation).  The  
ratio $r_T$ that is derived by using eq.~(\ref{r_C}), 
and the two consistency relations eq.~(\ref{consistency1}) 
and (\ref{consistency2}) are denoted by 
(i), (ii), (iii), respectively. Note that while the $r_T$ calculated 
numerically, (i), typically agreed with (ii), but it often differs from (iii).
\end{figcaption}
\end{center}
\end{figure}

\begin{figure}
\begin{center}
\singlefig{7cm}{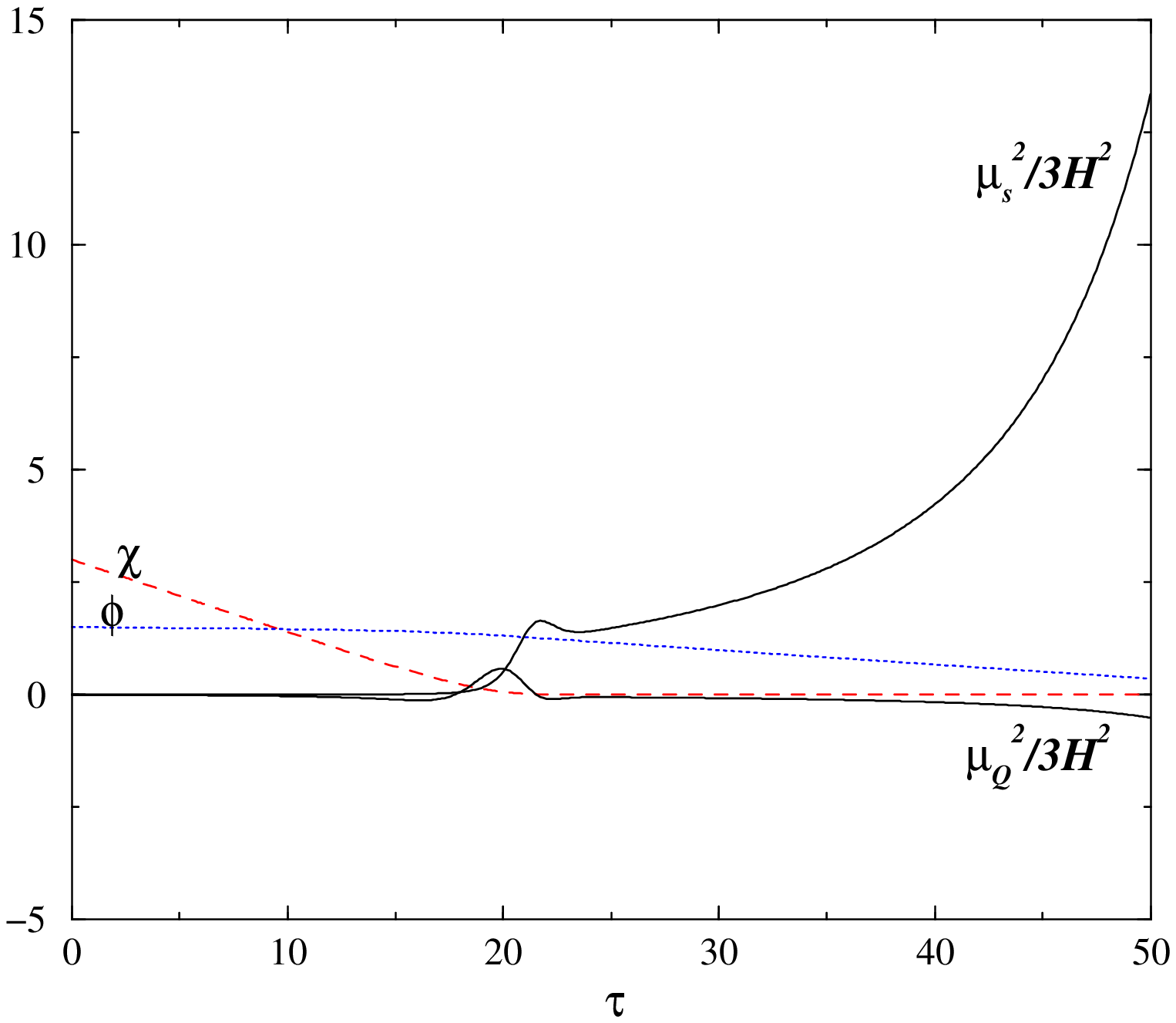}
\begin{figcaption}{qmphichi}{7cm}
The evolution of $\mu_Q^2/(3H^2)$ and $\mu_s^2/(3H^2)$ 
with $R = 5$, $m_\chi=1 \times 10^{-6} M_p$ and $g=0$.  
The initial conditions are 
chosen to be $\chi=3M_p$ and $\phi=1.5M_p$.  When the heavy field drops to 
the potential valley, a second phase of inflation begins, which is 
accompanied by the increase of $\mu_Q^2/(3H^2)$ and $\mu_s^2/(3H^2)$ 
The term $\mu_s^2/(3H^2)$ exhibits the growth by a factor 
of $5 \times 10^4$ by the end of inflation compared to its initial value.  
\end{figcaption}
\end{center}
\end{figure}
\begin{figure}
\begin{center}
\singlefig{7cm}{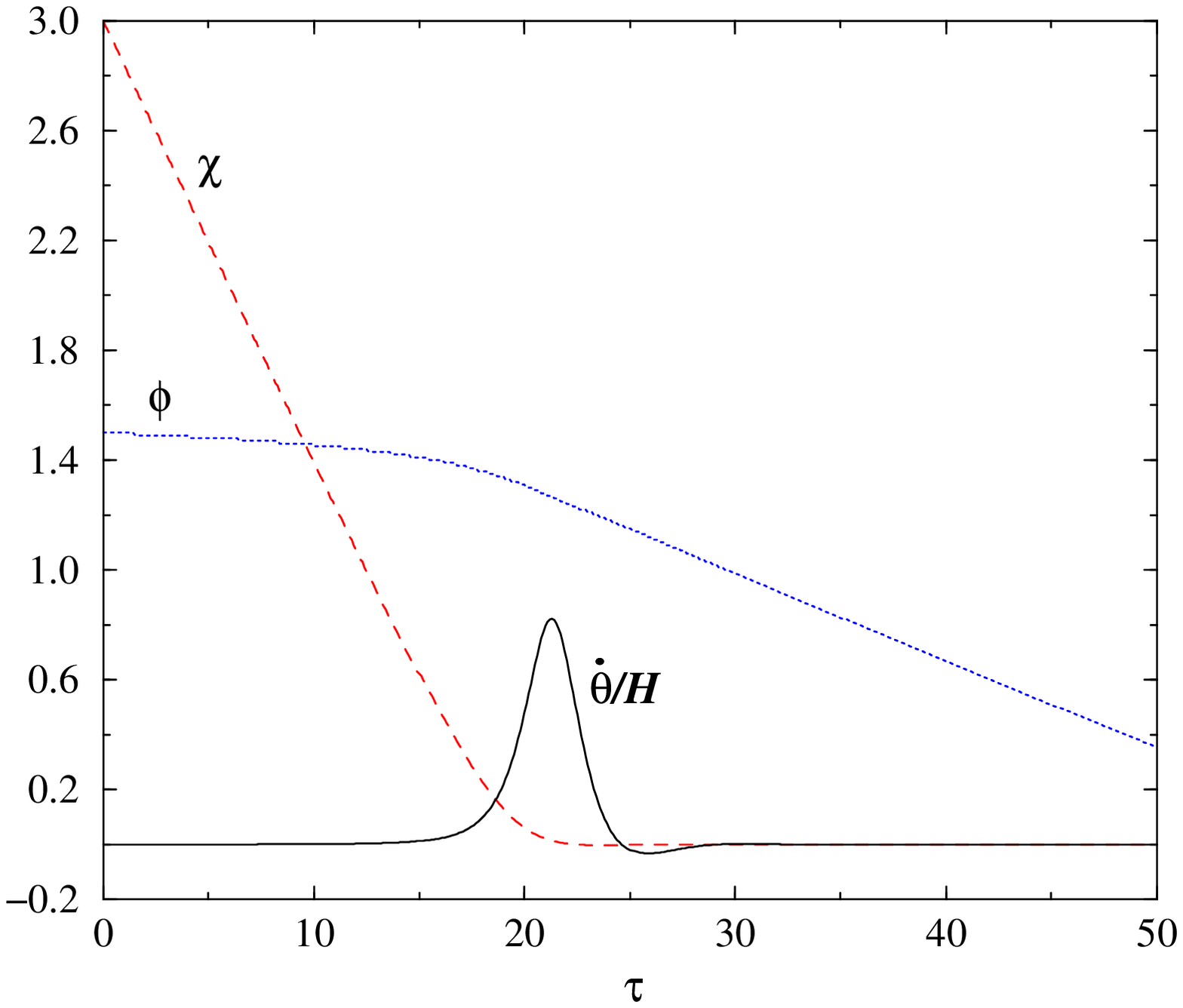}
\begin{figcaption}{phchtheta}{7cm}
The evolution of $\dot\theta/H$  with the same initial conditions as
in Fig.~\ref{qmphichi}.  When the heavy field drops to 
the potential valley, a second phase of inflation begins, which is 
accompanied by the increase of $\dot\theta/H$ from the initial value
$1.44 \times 
10^{-3} $ to its peak value $\dot\theta/H=0.8$ around the end of the first 
stage of inflation.
\end{figcaption}
\end{center}
\end{figure}

Two important quantities to determine the strength of the correlation 
are $R$ and $\tan \alpha_*$  around $N_H \simeq 60$
as seen from eq.~(\ref{x_g=0}). The e-folding of the second stage of
inflation, $N_0$, determines whether inflation is dominated by a heavy
or light scalar field around $N_H \simeq 60$ and also the strength of
the correlation on smaller scales.  Either of the scalar field masses,
$m_{\phi}$ or $m_{\chi}$, can be determined by the COBE normalization.
The relative ratio $R=m_\chi/m_\phi$ is important when we discuss the
correlation, $r_C$.  The correlation is strong around $1/R^2 \le \tan
\alpha_* \le 1$, 
whose lower bound is also determined by $R$.  If the precise observations 
in the future reveals the strength of the correlation around 
$50~\lsim~N_k~\lsim~63$, we will be able to constrain on two masses 
$m_{\phi}$ and $m_{\chi}$ (alternatively $R$ and $m_{\phi}$) together with 
the values of $\tan \alpha_*$ and $N_0$.

\subsection{The interacting case: $g \ne 0$}

Let us next consider the case where the coupling $g$ is taken into
account.  It was suggested by Linde and Mukhanov \cite{Linde:1996gt}
that inclusion of the coupling $g$ can lead to the blue spectrum of 
isocurvature perturbations.  Here we shall make detailed analysis
about the correlation of adiabatic and isocurvature perturbations. 

Let us first estimate the spectrum of isocurvature perturbations using the 
analytic estimates of Sec.~II.  
Although it has some errors due to the breakdown of slow-roll
approximations, it is still useful to make rough estimates
for the power spectrum.  The spectral index in eq.~(\ref{tiltS2}) is
estimated as 
\begin{eqnarray}
n_S-1=-2\epsilon_t+\frac{2\mu_s^2}{3H^2} \,.
\label{nSg}
\end{eqnarray}
Therefore it is important to consider the mass of the entropy field 
perturbation, $\mu_s$, relative to the Hubble rate, $H$.  Note that the 
term $-2\epsilon_t$ in the rhs of eq.~(\ref{nSg}) provides the slowly 
red-tiled spectrum.  If the mass square $\mu_s^2$ is larger than of order 
$H^2$, isocurvature perturbations are blue-tilted with $n_S>1$.  Making 
use 
of the slow-roll result (\ref{mu}), we find 
\begin{eqnarray}
\frac{2\mu_s^2}{3H^2}=
\frac{4(m_\chi^2+g^2\phi^2)(m_\phi^2+g^2\chi^2)
(m_\phi^2\phi^2+m_\chi^2\chi^2-2g^2\phi^2\chi^2)}
{\kappa^2(m_\phi^2\phi^2+m_\chi^2\chi^2+g^2\phi^2\chi^2)
\left\{(m_\phi^2+g^2\chi^2)^2\phi^2+(m_\chi^2+g^2\phi^2)^2
\chi^2 \right\}} \,.
\label{mu_S}
\end{eqnarray}

Let us first consider the case where $\mu_s^2$ is positive during the 
whole stage of double inflation, which corresponds to the condition, 
$m_\phi^2\phi^2+m_\chi^2\chi^2> 2g^2\phi^2\chi^2$.  When the heavy field 
$\chi$ rolls down to the valley $\chi=0$ at the first stage of inflation, 
we have $\mu_s^2 \simeq m_\chi^2+ g^2\phi^2$ and $3H^2 \simeq 4\pi 
m_\phi^2 \phi^2/M_p^2$.  
Then the mass square of $\delta s$ is given by 
\begin{eqnarray}
\mu_s^2 \simeq m_\chi^2+\beta H^2\,,~~~~{\rm with}~~~~ 
\beta=\frac{3g^2}{4\pi}\left(\frac{M_p}{m_\phi}\right)^2 \,.
\label{beta}
\end{eqnarray}
Note that in this case the entropy field perturbation $\delta s$
is almost the same as the heavy field perturbation $\delta \chi$.
If $\chi$ is quickly suppressed, we only need to consider $\delta \chi$, 
as in ref.~\cite{Linde:1996gt}, in order to discuss the spectrum of 
isocurvature 
perturbations.  When 
$\beta H^2$ is larger than $m_{\chi}^2$ during double inflation, we have 
$\mu_s^2 \simeq \beta H^2$ and 
\begin{eqnarray}
n_S-1 \simeq -2\epsilon_t+\frac23 \beta \,.
\label{nsg}
\end{eqnarray}

When $\beta$ is much larger than unity, this yields the 
blue-tilted spectrum, $n_S>1$.\footnote{When $\beta \gg 1$
the spectrum of isocurvature perturbations is highly blue-tilted.
This is actually the case for the preheating scenario 
where large-scale entropy field perturbations are strongly 
suppressed for the coupling $g$ required for strong preheating,
see refs.~\cite{preheat}.} Making use of this scenario, it is possible to 
obtain isocurvature perturbations that tend to grow toward smaller scales 
while adiabatic perturbations remain small on present horizon scales 
\cite{Linde:1996gt}.  If $\mu_s^2 \gg H^2$, then $\chi$ rolls down very 
rapidly to the local minimum of the potential valley ($\chi \to 0$), and 
$\dot{\theta}$ in eq.~(\ref{P_C}) exponentially decreases on smaller 
scales.  In this case the correlation between adiabatic and isocurvature 
perturbations tends to be very weak except for the scales where $\chi$ is 
not very small compared to $\phi$.  When $\dot{\theta}$ is negligible, the 
spectrum of curvature perturbations is practically no different from the 
single field result, $n_{\cal R}-1=-6\epsilon_\phi+2\eta_{\phi \phi}$ [see 
eq.~(\ref{tilt})].  In this case adiabatic perturbations can be nearly 
scale-invariant, while isocurvature perturbations are blue-tilted.

{}From eq.~(\ref{beta}) we find that the spectrum of isocurvature
perturbations can be blue-tilted for the coupling $g$ with 
$g~\gsim~m_{\phi}/M_p$.  In Fig.~\ref{gspectra} we plot the spectra of 
$P_R$, $P_S$ and $P_C$ for two cases with $\beta=0.01$ and $\beta=0.95$.  
Note that in these cases the model parameters are chosen so that $\mu_s^2$ 
is positive during the whole of double inflation.  When $\beta=0.01$, the 
spectrum of isocurvature perturbations is slightly blue-tilted, while for 
$\beta=0.95$ it is highly blue-tilted.  

The two spectra $P_{\cal R}$ and 
$P_C$ are not significantly modified by the presence of a coupling term 
$g$.  It can be understood that the correlation of adiabatic and isocurvature 
perturbations gets smaller as $\chi$ approaches the potential valley with 
decreasing $\dot{\theta}$.  As shown in Fig.~\ref{gconsist} the 
correlation 
$r_C$ tends to decrease more on smaller scales as we choose larger values 
of $\beta$.  When $\beta~\gsim~1$ we find that $r_C$ decreases rapidly on 
smaller scales, which is associated with the highly blue-tilted spectrum 
of isocurvature perturbations.  This is confirmed by the definition of $r_C$ 
in eq.~(\ref{r_C}) where only $P_S$ increases toward smaller scales.

{} From Fig.~\ref{gconsist} we find that the first consistency relation 
(\ref{consistency1}) exhibits fairly good agreement with $r_T$ obtained by 
eq.~(\ref{r_T}) except for larger scales, while the second one 
(\ref{consistency1}) does not.  This is caused by the violation of the 
slow-roll conditions at horizon crossing and also by the change of 
$\mu_Q^2/(3H^2)$, $\mu_s^2/(3H^2)$ and $\dot\theta/H$ during inflation.  
Since the correlation decreases toward smaller scales, the deviation from 
the numerical results tends to be weaker for smaller $N_k$ in ths case of 
the first consistency relation.  Since the second consistency relation is 
affected by the change of the mass terms after horizon crossing, it does 
not agree well with numerical results even on smaller scales.

\begin{figure}
\begin{center}
\singlefig{7cm}{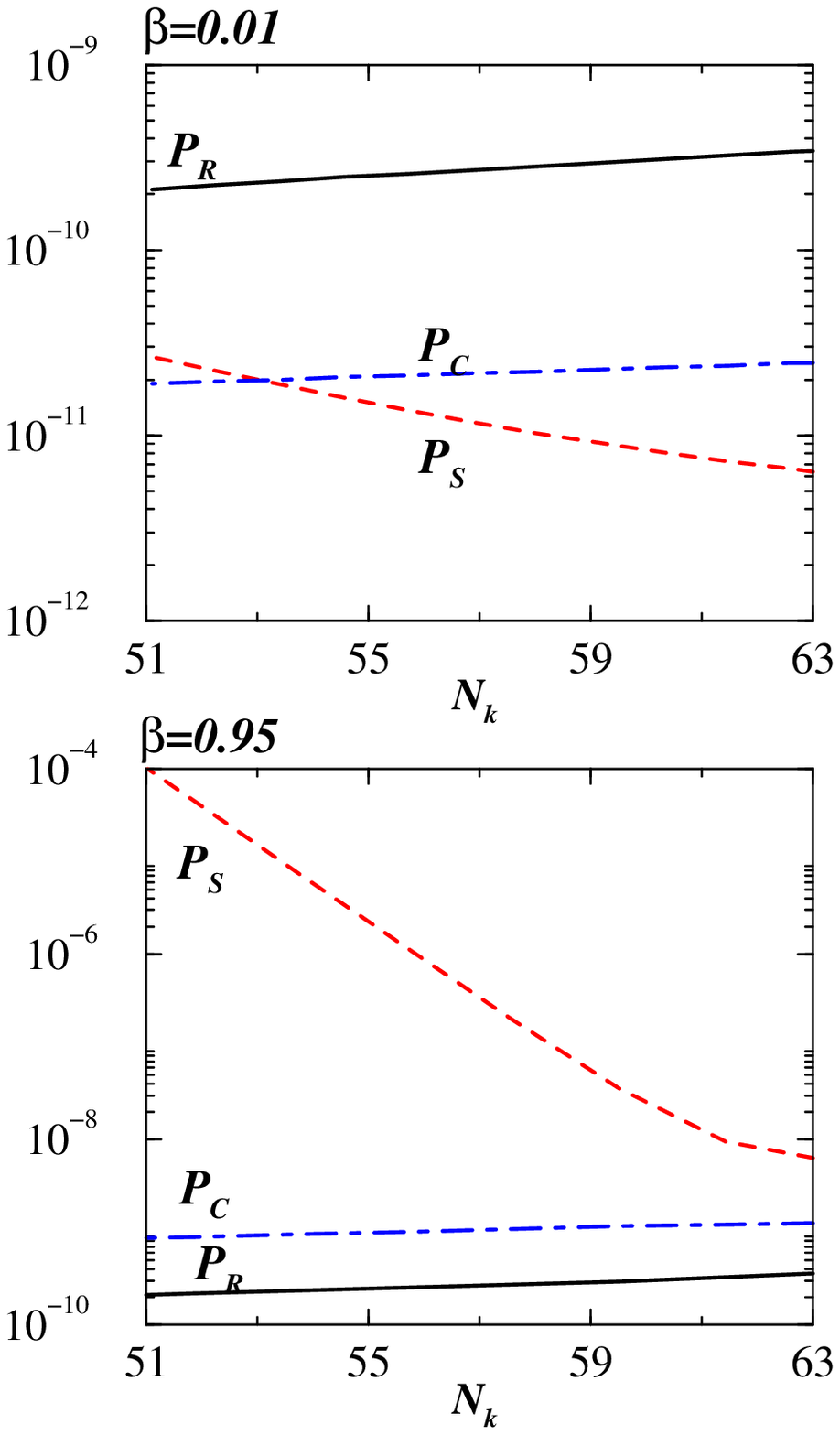}
\begin{figcaption}{gspectra}{8cm}
The power 
spectra $P_R$, $P_S$, and $P_C$ are shown for $\beta=0.01$ and 
$\beta=0.95$.
The model parameters are chosen to be $R=3$, $m_{\phi}=5.0 \times 
10^{-7}M_p$, and $\phi=3.2M_p$, $\chi=0.3M_p$ at $N_k=65$.  
\end{figcaption}
\end{center}
\end{figure}
\begin{figure}
\begin{center}
\singlefig{7cm}{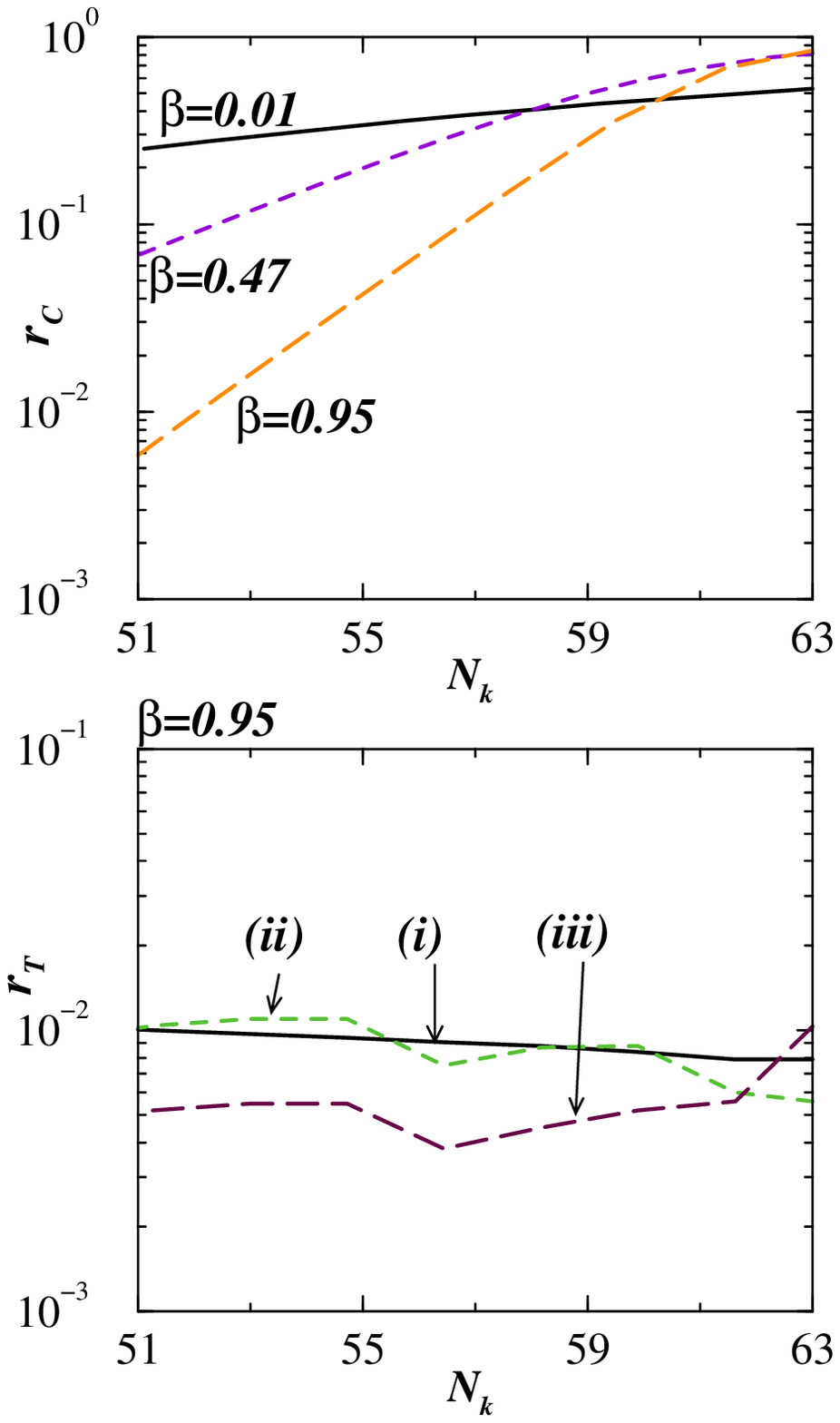}
\begin{figcaption}{gconsist}{8cm}
The correlation $r_C$ for $\beta=0.01, 0.47, 0.95$,
and the ratio $r_T$ which are derived by eqs.~(\ref{r_T}), 
(\ref{consistency1}) and (\ref{consistency2}), denoted by (i), (ii), and 
(iii), respectively.  
The model parameters are chosen to be the same as Fig.~\ref{gspectra}.
\end{figcaption}
\end{center}
\end{figure}

 Note that in Fig.~\ref{gconsist} 
the strength of the correlation $r_C$ increases for larger $\beta$ around 
the scale $N_H=60$.  Since the inclusion of the coupling $g$ provides the 
additional source term for $\dot{\theta}$ [see the $\eta_{\phi\chi}$ term 
in eq.~(\ref{dtheta})], this works to induce the larger correlation as 
long 
as $\chi$ is not strongly suppressed.  Making use of eq.~(\ref{dtheta}), 
we 
can easily show that the correlation is nonzero even for 
$R=1$.\footnote{We 
have $r_C=0$ for $R=1$ and $\phi=\chi$.} Fig.~\ref{rcg} indicates that the 
values of $r_C$ are increased around the region where the correlation is 
strong, by including the coupling $g$.
  
If the condition, $m_\phi^2\phi^2+m_\chi^2\chi^2<2g^2\phi^2\chi^2$, 
is satisfied at horizon crossing, the mass of $\delta s$ 
is {\it negative}.  So the spectrum of isocurvature 
perturbations produced is red tilted with a steeper slope than in the 
case of $g=0$.  Fig.~\ref{negative} corresponds to the case where the 
spectrum $P_S$ is  red-tilted for $57~\lsim~N_k~\lsim~63$ but begins to be 
blue-tilted for $N_k~\lsim~57$.  The negative mass of $\delta s$ leads to 
the red-tilted spectrum on large scales as expected.  When $\phi$ and 
$\chi$ 
are the same  order on these scales, the correlation $r_C$ can be close to 
unity [see the right panel of Fig.~\ref{negative}].  When the mass of 
$\delta s$ becomes positive and $\chi$ begins to decrease toward $\chi=0$, 
the situation is almost the same as discussed previously.  In this case we 
have highly blue-tilted spectrum for isocurvature perturbations with 
suppressed correlations ($r_C \ll 1$).  
 
\begin{figure}
\begin{center}
\singlefig{9cm}{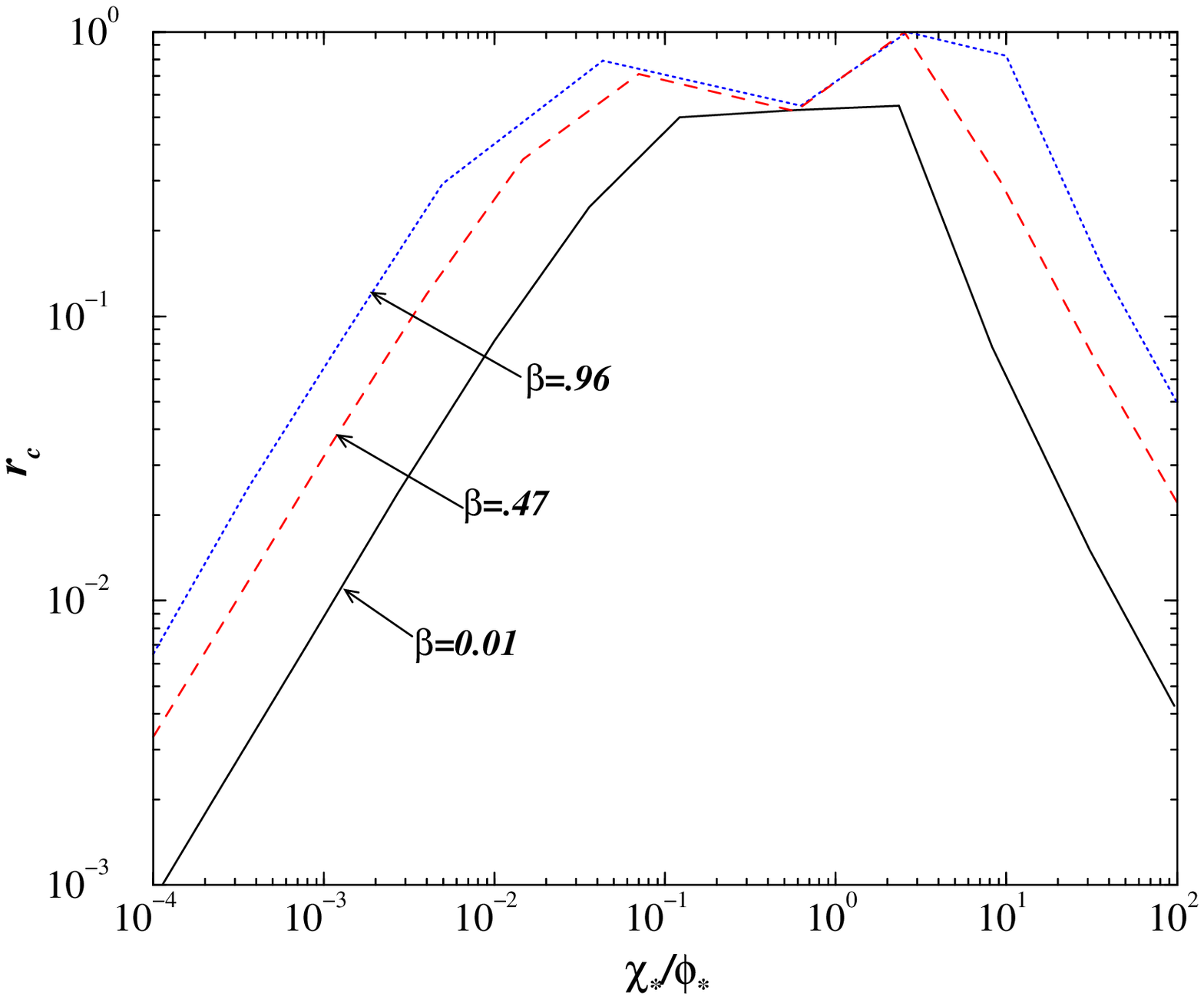}
\begin{figcaption}{rcg}{9cm}
The correlation $r_C$ as a function of $\chi_*/\phi_*$ for 
$\beta=0.01, 0.47, 0.95$ on the scale corresponding to 
$N_k=60$.  The model parameters are the 
same as in Fig.~\ref{gspectra}.
\end{figcaption}
\end{center}
\end{figure}

\begin{figure}
\begin{center}
\singlefig{7cm}{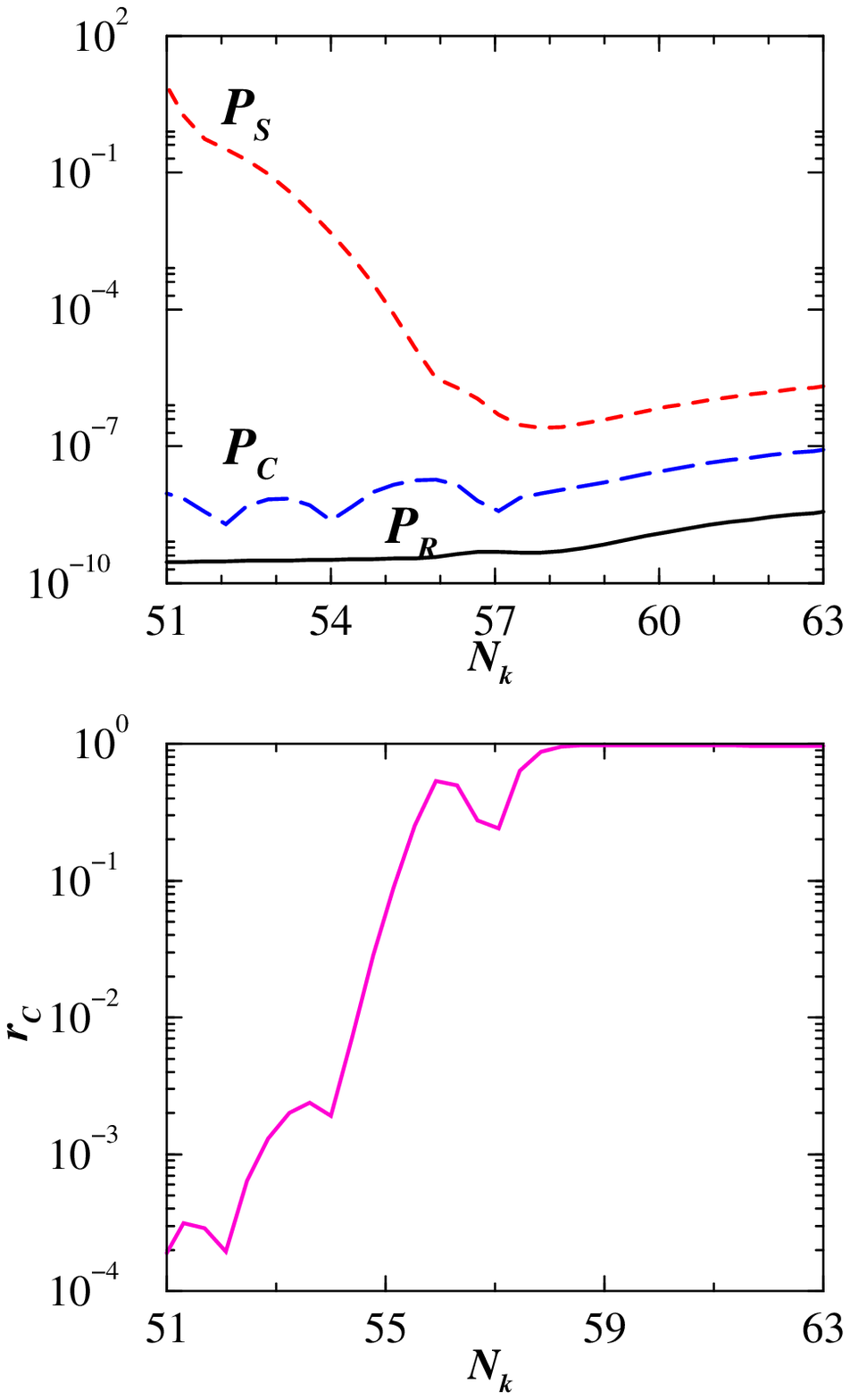}
\begin{figcaption}{negative}{7cm}
The power spectra $P_R$, $P_S$, $P_C$ (left) and the correlation 
$r_C$ for $R=3$, $m_{\phi}=2.0 
\times 10^{-7}M_p$ and $g=2.0 \times 10^{-6}$
(corresponding to $\beta=23.9$).
\end{figcaption}
\end{center}
\end{figure}

Unless $g$ is extremely small ($g \ll m_{\phi}/M_p$), 
then it is natural to have a stage of the negative 
$\mu_s^2$ during double inflation.  For example, when 
$g~\gsim~m_{\phi}/M_p$, it is easy to satisfy the condition, $\mu_s^2<0$, 
if $\chi$ is larger than the order of the Planck mass.  For the double 
inflationary scenario where inflation starts out with large initial values 
of $\phi$ and $\chi$ much greater than the Planck mass, the spectrum $P_S$ 
is highly red-tilted.  Nevertheless, when $g$ is large and $\beta \gg 
1$, $\chi$ decreases very rapidly toward $\chi=0$.  Therefore the 
blue-tilted spectrum of $P_S$ immediately appears once the mass of $\delta 
s$ becomes positive.

We have found that a variety of power spectra and correlations can be 
obtained, depending on the initial values of scalar fields and the 
parameters 
of the model.  In particular the inclusion of the coupling $g$ leads to an 
interesting power spectrum of isocurvature perturbations that 
tend to increase toward large scales (corresponding to $\mu_s^2<0$)
and also grow again toward smaller scales (corresponding to $\mu_s^2>0$).
If such a spectrum is supported from observations, it should be 
possible to constrain on the strength of the coupling $g$ and other model 
parameters by taking into account the information of the correlation $r_C$ 
as well.

There exist other models of double inflation 
which provide the $\beta H^2$ correction as in eq.~(\ref{beta}). 
One such model is a non-minimally coupled $\chi$
field with a minimally coupled  field $\phi$ \cite{Linde:1996gt}:
\begin{eqnarray}
V=\frac12 m_{\phi}^2 \phi^2+\frac12 m_{\chi}^2 \chi^2
+\frac12 \xi R \chi^2\,,
\label{nonminimal}
\end{eqnarray}
where $\xi $ is a non-minimal coupling between the scalar curvature
$R$ and the field $\chi$. In this model the spectrum of the isocurvature 
perturbations is red-tilted due to the amplification of $\delta \chi$
for negative $\xi$ \cite{Tsujikawa:2000wc,Starobinsky:2001xq}, 
while it is blue-tilted for positive $\xi$.
Although the decomposition into adiabatic and entropy ``fields''
is not as simple as in the case of minimally coupled fields 
discussed in Sec.~II, it would be of interest to extend our analysis
to this case.

\section{Double inflation motivated by supersymmetry}
\label{supergra}

We now come to the perhaps the most interesting of the models 
we have studied. In hybrid and supernatural inflationary 
models \cite{Linde:1993cn,Copeland:1994vg,RSG}, 
the symmetry breaking transition occurs in the presence 
of the second scalar field, 
$\chi$. The effective potential of the original hybrid 
inflation model is given by \cite{Linde:1993cn} 
\begin{eqnarray}
V= \frac{\lambda}{4} \left(\chi^2-\frac{M^2}
{\lambda}\right)^2 +\frac12 g^2 \phi^2 \chi^2+
\frac12 m^2\phi^2\,.
\label{hybrid}
\end{eqnarray}

This potential is closely related with those obtained in 
supersymmetric theories \cite{RSG} -\cite{Linde:1997sj}.  For example, 
consider the supersymmetric theory with a superpotential 
\begin{eqnarray}
W = S \left(\kappa_0 \vp \bar{\vp}-\mu^2 \right)\,,
\label{superpo}
\end{eqnarray}
which includes two superfields, $S$, $\varphi$, together with a conjugate 
pair, $\bar{\varphi}$. In the global supersymmetric limit 
($M_p \to \infty$), one obtains the following effective potential for two 
superfields $S$ and $\varphi$: 
\begin{eqnarray}
V=\left| \kappa_0 \varphi \bar{\varphi}-\mu^2 \right|^2+ \kappa_0^2 |S|^2 
\left( |\varphi|^2+|\bar{\varphi}|^2 \right) +D{\rm -terms}\,.
\label{effpo}
\end{eqnarray}
Note that this has a potential minimum at $|S|=0$, $\langle \varphi 
\rangle \langle \bar{\varphi} \rangle=\mu^2/\kappa_0$, 
$|\langle \varphi \rangle|= |\langle \bar{\varphi} \rangle|$.  
Making gauge and $R$-transformations in 
the $D$-flat direction, $|\langle \varphi \rangle|=|\langle \bar{\varphi} 
\rangle|$, the complex superfields, $S$, $\varphi$, $\bar{\varphi}$ can be 
replaced by real scalar fields, $\phi$ and $\chi$, as 
\begin{eqnarray}
S=\phi/\sqrt{2},~~~~\varphi=\bar{\varphi}=\chi/2.
\label{realsca}
\end{eqnarray}
Then the potential (\ref{effpo}) yields
\begin{eqnarray}
V=\frac{\kappa_0^2}{16}\left( \chi^2- \frac{4\mu^2}{\kappa_0} \right)^2+ 
\frac14 \kappa_0^2\phi^2\chi^2\,,
\label{effpo2}
\end{eqnarray}
where we neglected the $D$-terms.
The absolute minimum appears at $\phi=0$, $\chi=2\mu/\sqrt{\kappa_0}$.  
The potential (\ref{effpo2}) is exactly flat at the local 
minimum, $\chi=0$.  
Adding a mass term $\frac12 m^2\phi^2$ in 
Eq.~(\ref{effpo2}) results in the effective potential (\ref{hybrid}) 
with replacement, $\kappa_0^2/2=g^2=2\lambda$ and 
$\mu^2=M^2/(2\sqrt{\lambda})$.  
Therefore the supersymmetric version of the hybrid or double inflation 
corresponds to the case with $g^2/\lambda=2$.

Taking into account the supergravity correction 
gives rise to a slowly varying effective potential,
whose form is approximately given by 
$V \simeq \mu^4 \left[1+\phi^4/(8M_p^4)\right]$ 
\cite{Panagiotakopoulos:1997qd}.  
If one-loop radiative corrections are 
included, the total effective potential for 
$\phi>\sqrt{2}\mu/\sqrt{\kappa_0}$ involves logarithmic term, 
${\rm ln}\,\phi$, as well as the $\phi^4$ term \cite{Linde:1997sj}.  The corrections terms, 
$\phi^4$ or ${\rm ln}\,\phi$, can lead to an inflationary expansion of the 
universe for $\phi>\sqrt{2} \mu/\sqrt{\kappa_0}$.

Although these are different from the mass 
term $\frac12 m^2\phi^2$ in eq.~(\ref{hybrid}), the basic structures of 
the 
models motivated by supersymmetric theories are well described by the 
potential (\ref{hybrid}).  In particular, when we discuss the correlation 
between adiabatic and isocurvature perturbations, the crucial point is the 
evolution of scalar fields {\em after} the symmetry breaking phase 
rather than the early evolution at  $\phi>\sqrt{2}\mu/\sqrt{\kappa_0}$.  
Therefore we shall consider the model 
(\ref{hybrid}) in order to understand the basic properties of the 
correlations.  We are particularly interested 
in the supersymmetric case with $g^2/\lambda=2$.

\subsection{The condition for double inflation and the
background evolution } 

We shall first consider the evolution of the background and the condition 
for 
double inflation to take place (rather than just a single phase of 
inflation) 
for the model (\ref{hybrid}).  When $\phi$ is larger 
than $\phi_c \equiv M/g$, inflation takes place due to the slow-roll 
evolution of $\phi$.  Since the mass of $\chi$ is positive for 
$\phi>\phi_c$, the field $\chi$ rolls down to the potential valley at 
$\chi=0$.  Therefore the potential is approximately described as 
$V \simeq \frac{M^4}{4\lambda}+\frac12 m^2\phi^2$.  If the condition, 
$m^2\phi_c^2 \ll M^4/\lambda$, is satisfied, the Hubble constant at 
$\phi=\phi_c$ is given by $H \simeq H_0 \equiv 
\sqrt{2\pi/(3\lambda)}M^2/M_p$.  
Let us denote the masses of the two fields $\phi$ and $\chi$ relative to 
$H_0^2$ as $\gamma$ and $\delta$:
\begin{eqnarray}
\gamma \equiv \frac{m^2}{H_0^2}=
 \frac{3\lambda m^2 M_p^2}{2\pi M^4}, ~~~~\delta \equiv 
 \frac{g^2\phi^2-M^2}{H_0^2} =\frac{3\lambda}{2\pi}\left(\frac{M_p}{M} 
 \right)^2 (c^2-1)\,,
\label{albe}
\end{eqnarray}
where we set $\phi=c\phi_c$. $\gamma$ is required to be smaller than unity 
in order to lead to the first stage of inflation for $\phi>\phi_c$, 
thereby 
yielding 
\begin{eqnarray}
M^2~\gsim~mM_p \sqrt{\lambda}\,.
\label{del}
\end{eqnarray}

Whether the second stage of inflation occurs or not after $\phi$
drops below $\phi_c$ depends on the model parameters.  
If the ``water-fall'' condition, 
\begin{eqnarray}
M^3 \ll \lambda mM_{p}^2 \,,
\label{waterfall}
\end{eqnarray}
is satisfied, inflation soon comes to an end after the 
symmetry breaking.  This corresponds to the original version of the hybrid 
inflationary scenario where inflation ends due to the rapid rolling of the 
field $\chi$ \cite{Linde:1993cn}.

Combining eqs.~(\ref{del}) and (\ref{waterfall}), one has $M \gg m$ and
\begin{eqnarray}
\delta \gg \frac{M}{m} (c^2-1) \gg c^2-1 \,.
\label{delap}
\end{eqnarray}
This means that the classical field 
$\chi$ is strongly suppressed for $\phi>\phi_c$ ($\chi \propto a^{-3/2}$).
Since inflation typically starts when the value of $c^2-1$ is of order 
unity or much larger than unity, it is inevitable to avoid the suppression 
of 
$\chi$ when the water-fall condition is satisfied.
Note that $\delta$ changes sign after the symmetry breaking.  The field 
$\chi$ and its large-scale fluctuations are amplified by the tachyonic 
instability associated with negative $\chi$ 
mass \cite{Boyanovsky:1997xt,Cormier:1998nt,Tsujikawa,Felder:2000hj}.

Although the growth is strong for large-scale modes ($k \to 0$), 
the size of these fluctuations is 
vanishingly small at the beginning of the tachyonic 
instability due to their exponential suppression for $\phi>\phi_c$.  
Therefore the small-scale modes that are not significantly suppressed for 
$\phi>\phi_c$ provide the larger contribution to the total variance 
$\langle
\chi^2 \rangle$ of  $\chi$ rather than the large-scale modes.

The condition for the second stage of inflation to occur
is characterised by $|\delta| \ll 1$, namely
\begin{eqnarray}
M^2 \gg \lambda M_{p}^2 \,.
\label{double}
\end{eqnarray}
In this case the field $\chi$ and its large-scale perturbation are free 
from 
the inflationary suppression for $\phi>\phi_c$, unless inflation starts 
out 
with very large values of $\phi$ satisfying $c \gg 1$.
Note that one has $m^2/M^2 \ll g^2/\lambda$ under the 
condition that the first stage of inflation is driven by 
the Hubble constant, $H_0$ (namely, $m^2\phi_c^2 \ll 
M^4/\lambda$).  

Therefore one has $M \gg m$ for 
$g^2/\lambda ={\cal O}(1)$.  Combining this relation with 
eq.~(\ref{double}) gives $M^3 \gg \lambda mM_{p}^2$,
which means that the water-fall condition (\ref{waterfall}) is violated.  
In 
this case the evolution of the field $\chi$ is sufficiently slow so that 
the 
second stage of inflation occurs after the symmetry breaking.

\begin{figure}
\begin{center}
\singlefig{10cm}{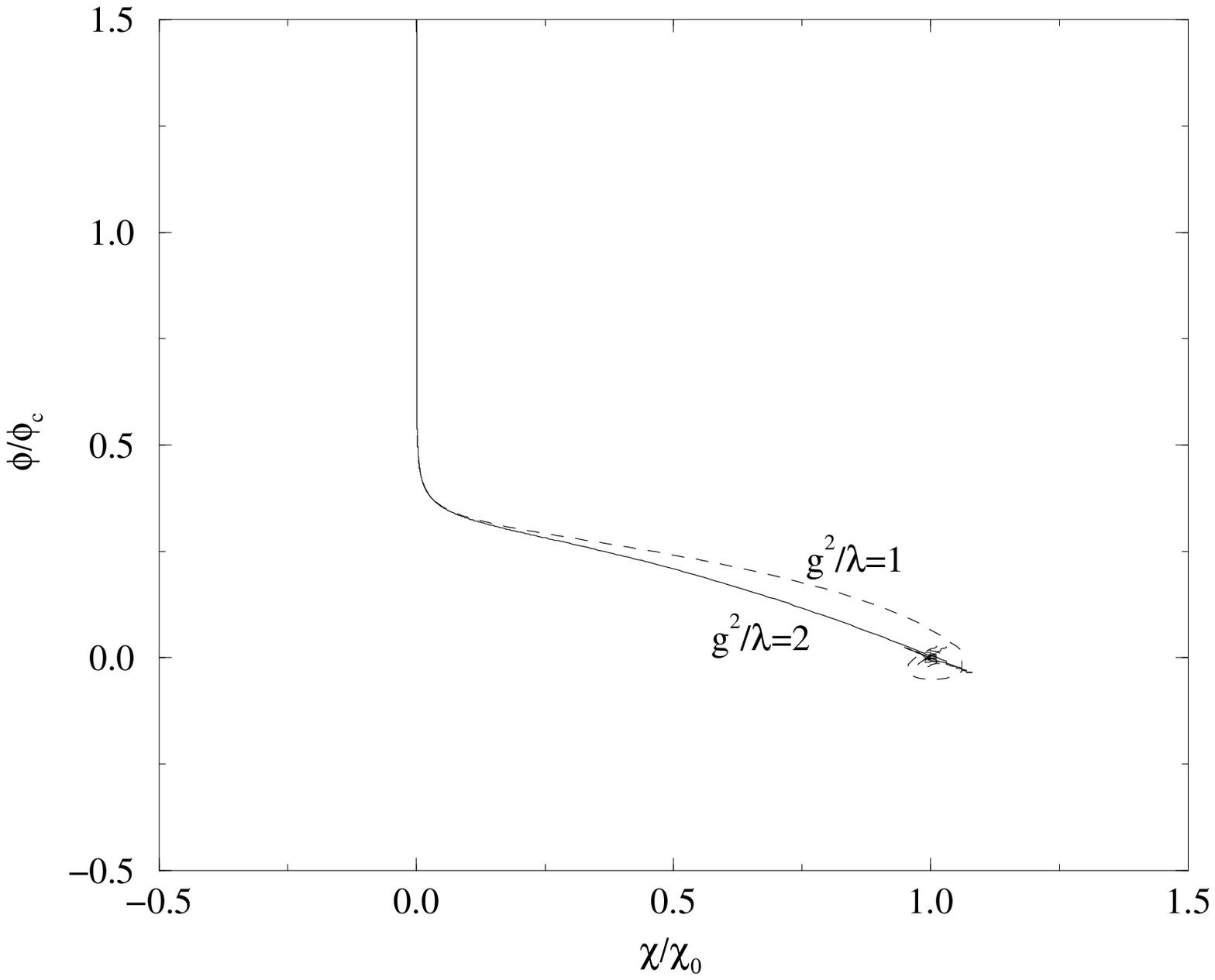}
\begin{figcaption}{fieldspace}{10cm}
The trajectory of two scalar fields in the plane ($\phi/\phi_c, 
\chi/\chi_0$).
Model parameters are chosen to be $M=7.0 \times 10^{-7}M_p$, 
$m=2.0 \times 10^{-7}M_p$ with initial scalar fields 
$\phi_i=1.5\phi_c$ and $\chi_i=10^{-3}\chi_0$.
We show two cases of $g^2/\lambda=1$ and $g^2/\lambda=2$  
with $\lambda=10^{-12}$. 
The trajectories are curved in field space, which means that 
$\dot{\theta} \ne 0$.  
\end{figcaption}
\end{center}
\end{figure}

Let us consider the evolution of the background for 
$g^2/\lambda={\cal O}(1)$.  The number of e-folds 
during the first stage of inflation is described as 
\begin{eqnarray}
N_1 \simeq \kappa^2 \int_{\phi_c}^{\phi_i} 
\frac{V}{V'}\,d\phi \simeq \frac{2\pi M^4}
{\lambda m^2M_p^2}\,{\rm ln} \frac{\phi_i}{\phi_c} \,,
\label{N1}
\end{eqnarray}
where we used $V \simeq \frac{M^4}{4\lambda}+\frac12 m^2\phi^2$ for 
$\phi>\phi_c$.  Here $\phi_i$ is the value of $\phi$ at the beginning of 
double inflation.  Note that we have $N_1 \gg 1$ under the condition of 
eq.~(\ref{del}) [i.e., $\gamma \ll 1$].  Similarly the number of e-folds 
after the symmetry breaking is approximately expressed as 
\begin{eqnarray}
N_2 \simeq \kappa^2 \int_{\chi_0}^{\chi_c} \frac{V}{V'}\,d\chi \simeq 
\frac{2\pi M^2}{\lambda M_p^2}\,{\rm ln} \frac{\chi_0}{\chi_c} \,,
\label{N2}
\end{eqnarray}
where we used $V \simeq \frac{\lambda}{4}\left(\chi^2-\frac{M^2} 
{\lambda}\right)^2$.  Here $\chi_0=M/\sqrt{\lambda}$ and $\chi_c$ is the 
value of $\chi$ at $\phi=\phi_c$.  Again $N_2 \gg 1$ is satisfied under 
the 
condition of eq.~(\ref{double}).  We are interested in the double 
inflationary scenario where the total amount of e-folds, $N_T=N_1+N_2$, 
exceeds $N_H=60$.

When $g^2/\lambda={\cal O}(1)$, the critical value $\phi_c=M/g$
and the potential minimum $\chi_0=M/\sqrt{\lambda}$ are of the 
same order.  Two fundamental masses around the potential minimum
are characterised by $m_{\phi} \equiv (g/\sqrt{\lambda})M$ and 
$m_{\chi} \equiv \sqrt{2}M$.  Therefore these masses are also comparable 
when $g^2/\lambda={\cal O}(1)$.  In particular in the supersymmetric case 
with $g^2/\lambda=2$, the two masses are completely equal.  

In this case the 
trajectory of the two scalar fields after the symmetry breaking is close 
to a 
straight line in the ($\phi/\phi_c, \chi/\chi_0$) plane if the velocities 
of $\phi$ and $\chi$ are sufficiently small at the bifurcation point, 
$\phi=\phi_c$ \cite{hybridpre2}.  However, since $\dot{\phi}$ is non-zero 
because of the non slow-roll evolution around $\phi=\phi_c$, the 
trajectory 
is not strictly described by a straight line after the symmetry breaking.  
In 
fact this behaviour can be found in our numerical simulation in 
Fig.~\ref{fieldspace}.  When $g^2/\lambda ={\cal O}(1)$ and $g^2/\lambda 
\ne 2$ the two scalar fields exhibit chaotic behaviour as shown in 
refs.~\cite{hybridpre2,Easther:1997hm,hybridpre1}.  The trajectory in the 
$g^2/\lambda=1$ case is illustrated in 
Fig.~\ref{fieldspace}.\footnote{Note 
that the amplitude of the two scalar fields can be higher as in 
ref.~\cite{hybridpre1,hybridpre2} by changing model parameters.} Since the 
trajectory of the two scalar fields is generally curved, this leads to the 
variation of $\theta$ in field space ($\dot{\theta} \ne 0$), thereby 
generating a correlation of perturbations for $\phi<\phi_c$.  Note that 
in the case of $g^2/\lambda \ll 1$ or $g^2/\lambda \gg 1$, $m_{\phi}$ and 
$m_{\chi}$ as well as $\phi_c$ and $\chi_0$ take quite different values.  
We will not consider such cases in this work, since we are interested in 
the double inflation motivated by supersymmetric theories.

\subsection{Perturbations}

Let us next analyze the perturbations and correlations in the
double inflation model with potential (\ref{hybrid}).  
When the field $\phi$ evolves slowly along the potential valley 
with $\chi=0$ before the symmetry breaking, the spectral index
of the curvature perturbation generated in the first stage of double 
inflation can 
be estimated by eq.~(\ref{tilt}), as 
\begin{eqnarray}
n_{\cal R}-1 \simeq -6\epsilon_{\phi}+2\eta_{\phi \phi}
\simeq \frac 23\,\gamma \left(1-\frac{3m^2\phi^2}{V}\right) \,,
\label{nRs}
\end{eqnarray}
where $\gamma$ is defined by eq.~(\ref{albe}).
When the condition, $m^2\phi^2 \ll V \simeq M^4/(4\lambda)$, 
holds as is the case with the original hybrid inflation scenario 
\cite{Linde:1993cn}, one has the blue-tilted spectrum with 
$n_{\cal R}-1 \simeq \frac 23\,\gamma >0$. 
 Similarly the spectral index of the 
isocurvature perturbation generated for $\phi>\phi_c$ is given as 
\begin{eqnarray}
n_S-1 \simeq -2\epsilon_{\phi}+2\eta_{\chi \chi} \simeq \frac 
23\,\delta-\gamma \frac{m^2\phi^2}{3V} \,,
\label{nSs}
\end{eqnarray}
where we used eq.~(\ref{tiltS2}).
Therefore when the condition, 
$\frac23 \delta>\gamma \frac{m^2\phi^2}{3V}$, is satisfied, 
the isocurvature perturbation is also blue-tilted.  
Note that the spectral index of the 
correlation $P_C$ is similar to that of $P_S$ except for the last term in 
eq.~(\ref{tiltS}) that is of order $1/N_k \ll 1$ when $|\zeta_k N_k| \ll 
1$.

The spectral indices in eqs.~(\ref{nRs}) and (\ref{nSs}) can be modified 
in 
the presence of the tachyonic instability region with $\phi<\phi_c$.  
After 
the symmetry breaking, the field perturbation $\delta \chi$ begins to be 
amplified due to the negative $\chi$ mass in eq.~(\ref{albe}) with $c<1$.  
This growth is accompanied with the amplification of the entropy field 
perturbation $\delta s$ for small $k$ modes, which stimulates the 
enhancement of large-scale curvature perturbations by the relation 
(\ref{dotR}) [see Fig.~\ref{sugraevo}].  

\begin{figure}
\begin{center}
\singlefig{11cm}{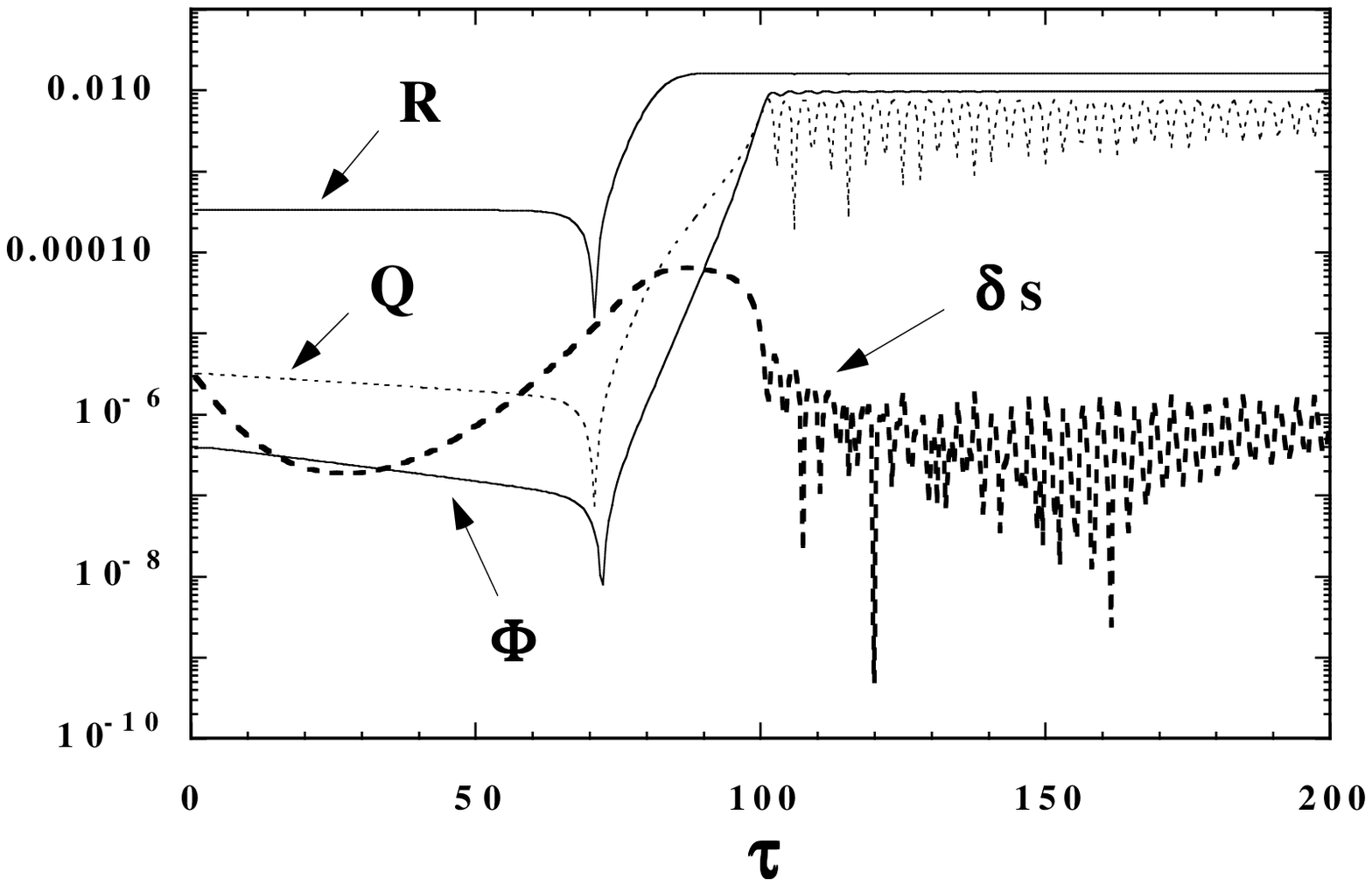}
\begin{figcaption}{sugraevo}{11cm}
The evolution of ${\cal R}$, $\Phi$, $\delta s$ and $Q$
for a mode which left the horizon before 60 e-foldings
from the end of the double inflation.
Note that we showed ${\cal R}=\sqrt{P_{\cal R}}$, e.t.c. 
Model parameters are 
$g^2/\lambda=2$, $g=1.5 \times 10^{-10}$, $M=5.0 \times 10^{-6}$ and 
$m=0.2M$ with initial conditions$, \phi=1.34\phi_c$ and $N=10^{-3} \chi_0$.  
${\cal R}$ and $\Phi$ are amplified due to the tachyonic growth of $\delta 
s$ and $Q$ during the second stage of inflation.
\end{figcaption}
\end{center}
\end{figure}

As shown in Fig.~\ref{qmuth}, 
$|\dot{\theta}/H|$ decreases during the first stage of inflation, but 
begins to increase after the symmetry breaking.  This can lead to the 
strong correlation between adiabatic and isocurvature perturbations.  In 
fact once $\delta s$ and $|\dot{\theta}/H|$ grow sufficiently, this works 
as source terms for $Q$ in the r.h.s. of eq.~(\ref{SMeq}), thereby 
stimulating the growth of $\Phi$ through the relation (\ref{Phisource}).  
This behaviour is clearly seen in the numerical simulation of 
Fig.~\ref{sugraevo}.

\begin{figure}
\begin{center}
\singlefig{11cm}{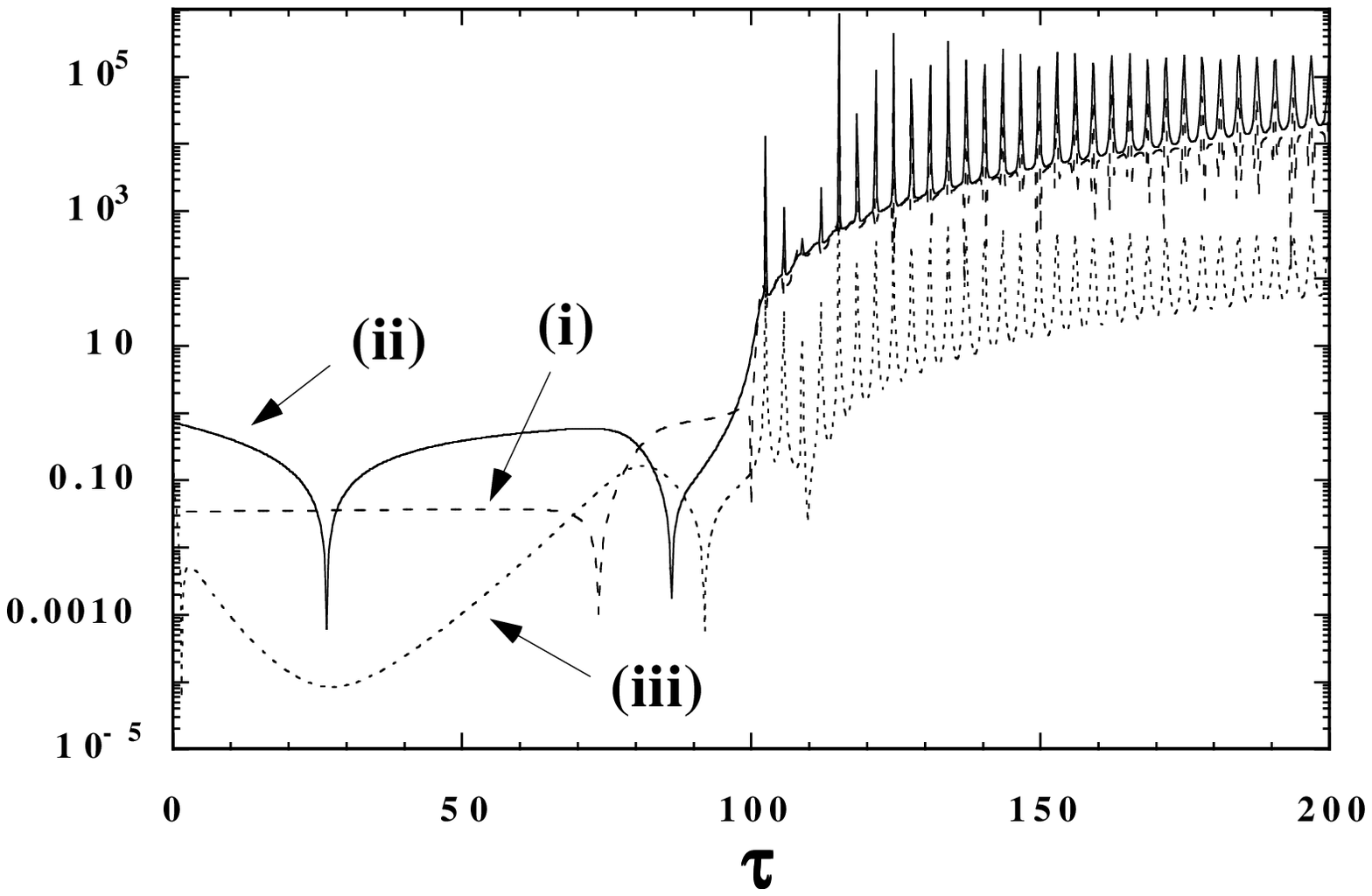}
\begin{figcaption}{qmuth}{11cm}
The evolution of (i) $|\mu_Q^2/(3H^2)|$, (ii) $|\mu_s^2/(3H^2)|$ 
and (iii) $|\dot\theta/H|$ for $g^2/\lambda=2$, 
$g=1.5 \times 10^{-10}$, $M=5.0 \times 10^{-6}$ and 
$m=0.2M$ with initial conditions$, 
\phi=1.34\phi_c$ and $\chi=10^{-3} \chi_0$.  
Although we showed the absolute values of these quantities, 
it happens that these take negative values in the tachyonic instability 
region.
\end{figcaption}
\end{center}
\end{figure}

Let us consider the spectra of perturbations at the end of 
the double inflation.  In Fig.~\ref{sugracon} we show the spectra $P_R$, 
$P_S$, and $P_C$ around the scale $N_H=60$ for three different cases.  
The case (a) corresponds to the one with $\gamma \simeq 0.08 \ll 1$ and 
$\delta \simeq c^2-1 \simeq 1$ around $N_k \sim 60$, in which case from 
eqs.~(\ref{nRs}) and (\ref{nSs}) one has a slight blue-tilt for $P_{\cal 
R}$ and a rather steep blue-tilt for $P_S$ at the end of the {\it 
first} stage of the double inflation.  

In fact we have numerically checked 
that such spectra are generated before symmetry breaking.  However 
these are different from the final spectra obtained at the end of 
double inflation.  Since the strong conversion between adiabatic and 
isocurvature perturbations occurs during the tachyonic instability region, 
the final spectrum of curvature perturbations is affected by the steep 
blue-tilted spectrum of isocurvature perturbations.  Therefore the final 
$P_{\cal R}$ exhibits the steeper blue-tilted spectrum than predicted by 
eq.~(\ref{nRs}).  

This tells us that the correlation between adiabatic 
and isocurvature perturbations is important to correctly estimate the 
final spectra.  The slow-roll results (\ref{nRs}) and (\ref{nSs}) 
typically 
show limitations when the correlation is strong.  Note that in 
Fig.~\ref{sugracon} all spectra $P_R$, $P_S$, and $P_C$ in the case (a) 
exhibit almost the same blue spectral indices due to the strong 
correlation.

Although the case (a) corresponds to the one with rather steep
blue-tilted spectra, one can obtain nearly scale-invariant 
spectra by choosing small values of $\gamma$ and $\delta$
relative to unity.  For example the case (b) in 
Fig.~\ref{sugracon} corresponds to the one with 
$\gamma \simeq 0.04 \ll 1$ and $\delta \simeq 
0.6(c^2-1)~\lsim~0.2$ for $N_k~\lsim~63$.  

In this case both the adiabatic 
and isocurvature spectra generated for $\phi>\phi_c$ are slightly 
blue-tilted, as predicted by eqs.~(\ref{nRs}) and (\ref{nSs}).  The 
conversion of perturbations occurs after the symmetry breaking as well, 
but the spectral indices are mostly inherited by the end of double inflation 
because both $P_{\cal R}$ and $P_S$ have similar small spectral indices at 
$\phi=\phi_c$.  As shown in Fig.~\ref{sugracon} all of $P_R$, $P_S$, and 
$P_C$ exhibit a slightly blue-tilted spectra at the end of double 
inflation.

\begin{figure}
\begin{center}
\singlefig{6.5cm}{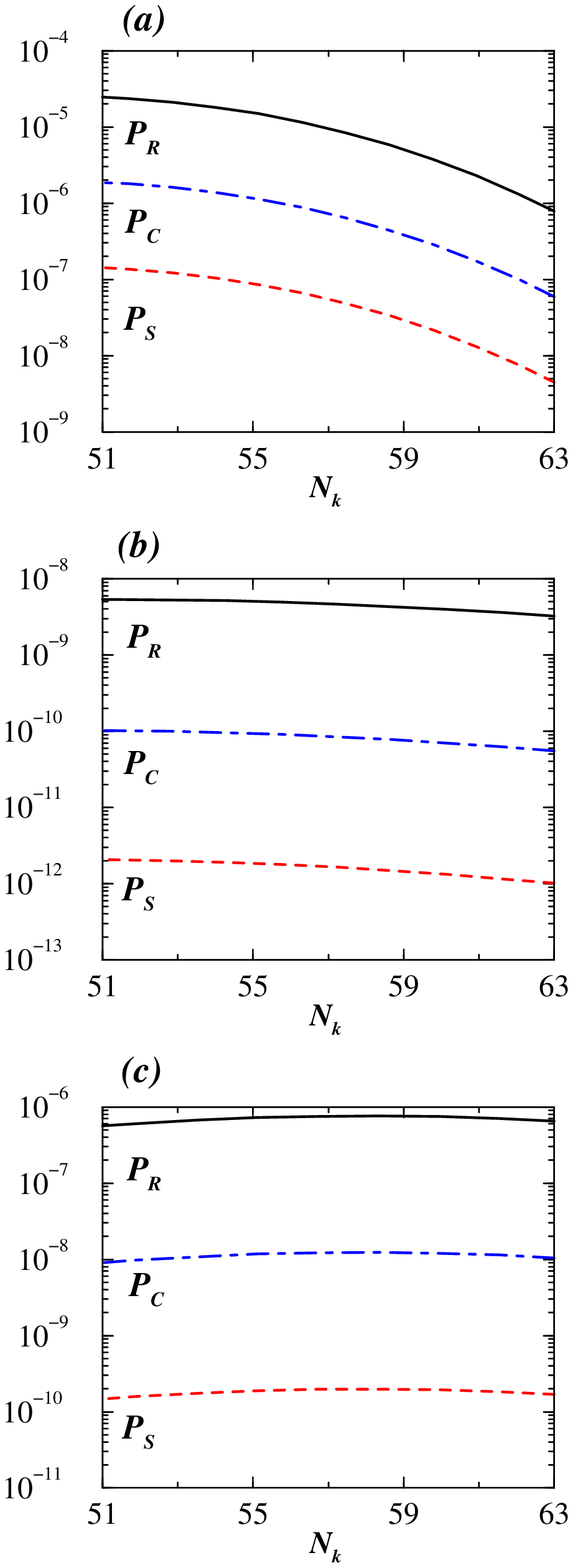}
\begin{figcaption}{sugracon}{6.5cm}
The power spectra $P_R$, $P_S$ and $P_C$
for $g^2/\lambda=2$. Each case corresponds to 
(a) $M=7.0 \times 10^{-7}M_p$, $\lambda=1.0 \times 10^{-12}$,
$m=2.0 \times 10^{-7}M_p$, $\phi_i=1.47\phi_c$, 
$\chi_i=1.0 \times 10^{-3}\chi_0$, 
(b) $M=8.5 \times 10^{-7}M_p$, $\lambda=9.0 
\times 10^{-13}$, $m=2.0 \times 10^{-7}M_p$, 
$\phi_i=1.22\phi_c$, $\chi_i=5.0 \times 10^{-2}\chi_0$,
(c) $M=8.1 \times 10^{-7}M_p$, $\lambda=1.0 \times 10^{-12}$, 
$m=2.0 \times 10^{-7}M_p$, $\phi_i=1.11\phi_c$, 
$\chi_i=1.0 \times 10^{-3}\chi_0$.
\end{figcaption}
\end{center}
\end{figure}

\begin{figure}
\begin{center}
\singlefig{6.5cm}{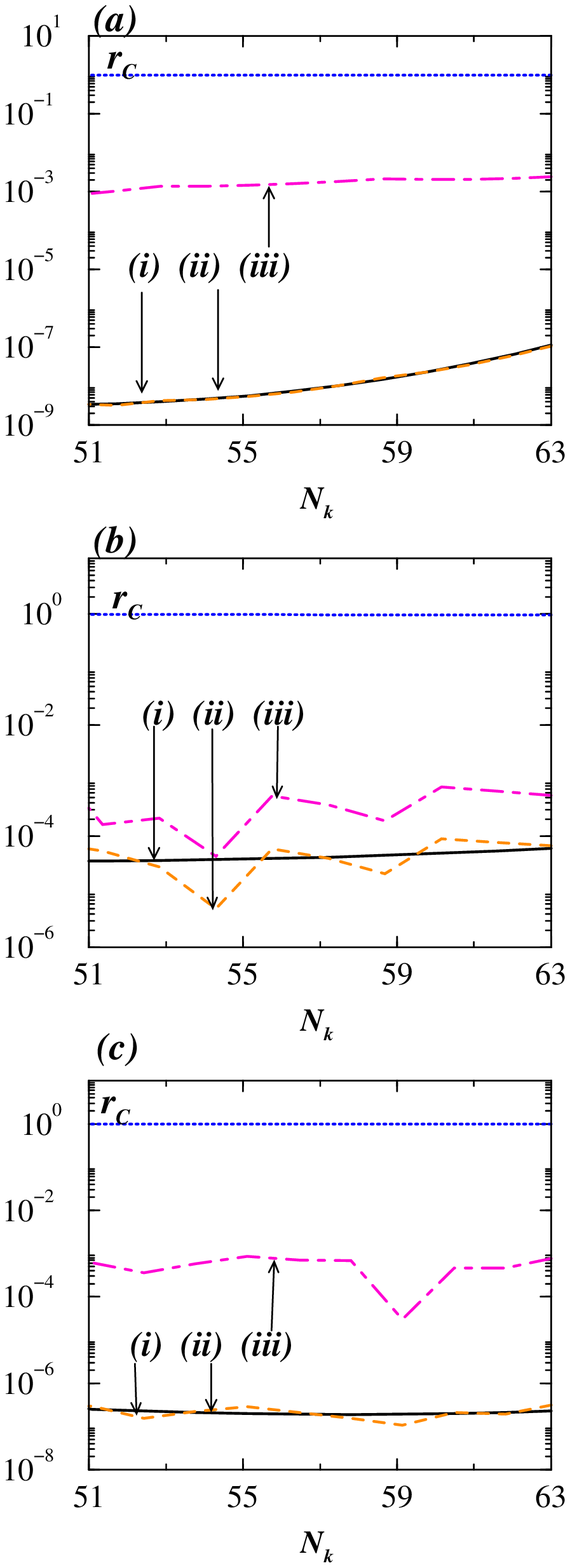}
\begin{figcaption}{suconsist}{6.5cm}
The correlation $r_C$ and the ratio $r_T$ that 
are derived by using eq.~(\ref{r_C}), and two consistency relations 
(\ref{consistency1}) and (\ref{consistency2}), which are denoted by (i), 
(ii), (iii), respectively.  We show the cases (i), (ii) and (iii)
by solid curves, dashed curves and dotted-dashed curves, respectively.  Note 
that in the case (\ref{consistency2}) we have taken the absolute value of 
$r_T$.  The initial condistions for the three cases are the same as in 
Fig.~\ref{sugracon}.
\end{figcaption}
\end{center}
\end{figure}

One may consider that the tachyonic growth of large-scale 
perturbations may lead to the red-tilted spectra.
In the cases (a) and (b) all modes shown in Fig.~\ref{sugracon} 
(corresponding to $51~\lsim~N_k~\lsim~63$) already left far outside of 
the horizon when the field reaches $\phi=\phi_c$.  
Since the physical momenta 
satisfy $k/a \ll H$ for these all modes, the tachyonic growth rate of 
perturbations is practically the same for the modes corresponding to 
$51~\lsim~N_k~\lsim~63$.  Therefore in the cases (a) and (b) the presence 
of the tachyonic region does not yield the red-tilted spectra.  

However, if 
the duration in the first stage of inflation is short, it is possible to 
obtain the red-tilted spectrum on smaller scales.  For example, in the 
case (c) illustrated in Fig.~\ref{sugracon}, the e-folds during the first stage 
of inflation are $N_1 \sim 7.5$ (the total e-folds are $N \sim 65$).  The 
modes corresponding to $N_k~\gsim~58$ crossed the horizon before the field 
reaches to the point $\phi=\phi_c$.  For these modes the spectra of 
perturbations are blue-tilted as are the cases of (a) and (b).  

In contrast, the smaller-scale modes with $N_k~\lsim~58$ crossed 
the horizon after the 
symmetry breaking, in which case one has the red-tilted spectrum due to 
the negative $\chi$ mass [see Fig.~\ref{sugracon}].  The case (c) corresponds 
to the slightly red-tilt spectra with $|\delta| \ll 1$.  If the values of 
$|\delta|$ are increased, we have steeper negative tilts than shown in 
Fig.~\ref{sugracon}.  It is very interesting that such a variety of 
spectra can be obtained by different choices of model parameters 
and initial conditions.

In Fig.~\ref{qmuth} we find that the absolute values of the mass 
$\mu_s^2/(3H^2)$ and $\dot{\theta}/H$ change during the 
double inflation, while the variation of 
$\mu_Q^2/(3H^2)$ is small.  
In addition, although the mass $\mu_s^2/(3H^2)$ is positive
initially, it changes sign after the symmetry breaking.
Therefore, to use the ``frozen" 
positive mass in eq.~(\ref{g}) is not 
typically valid, thereby leading to the errors in the final consistency 
relations.  And while the correlation is suppressed for $\phi>\phi_c$, the 
tachyonic growth of the fluctuation $\delta\chi$ yields the strong 
correlation after the symmetry breaking.  

Numerically we found that the correlation ratio, $r_C$, is very close to 
unity at the end of double inflation (see Fig.~\ref{sugracon}).  This is 
associated with the enhancement of ${\cal R}$ and $\Phi$ shown in 
Fig.~\ref{sugraevo}.  In Fig.~\ref{suconsist} the first consistency 
relation shows good agreement with the numerical results in the cases (a) 
and (c), while the case (b) is not so good.  In the cases (a) and (c) we 
chose the initial value $\chi_i=10^{-3}\chi_0$, while the case (b) 
corresponds to $\chi_i=0.04\chi_0$.  In the former cases one has 
$\dot{\theta}/H$ of order $0.001$ around the scale $N_k \sim 60$, but 
$\dot{\theta}/H$ is larger by more than one order of magnitude in the 
latter case.  The correlation is negligible at horizon crossing in the 
cases (a) and (c), but in case (b) it is not.  This is the main reason of 
the deviation from the first consistency relation in the case (b).  In fact 
we have numerically checked that the first consistency relation tends to 
agree with the numerical results as we decrease the initial $\chi$ (i.e., 
smaller $\dot{\theta}/H$).
Note that $r_C$ grows close to unity during the second stage of inflation, 
whose behavior is almost independent on the value of $r_C$ at horizon crossing.

Our numerical simulations show that the second consistency relation 
does not agree with the one obtained by the definition (\ref{r_T}) [see 
Fig.~\ref{suconsist}].
In particular, although $r_T$ is positive-definite in 
eq.~(\ref{r_T}), negative values of $r_T$ appear when we use 
eq.~(\ref{consistency2}), implying strong deviations from the second 
consistency relation [Note that in Fig.~\ref{suconsist} we showed the 
absolute values of $r_T$].  Again this is mainly due to the violation of 
the assumption of the constant masses and $\dot{\theta}/H$ during the 
tachyonic instability region.

Notice also that if we use the slow-roll expression for $x$ 
in eq.~(\ref{xslow})
this does not provide the correct value of the correlation $r_C$.
In the case (a) of Fig.~\ref{suconsist}, for example, we have 
$(\dot{\theta}/H)_k \sim 0.001$ and $\zeta_k \sim 0.37$ around
$N_k=60$.  Therefore eq.~(\ref{xslow}) leads to $x \sim 0.005$ and 
$r_C \sim 0.005 \ll 1$.  This is significantly different from the 
numerical value of $r_C$ close to unity.  We have to integrate 
the $\dot{\theta}/H$ term from the horizon crossing to the end of 
inflation in order to correctly estimate the final value of $r_C$.
Note that when we evaluate $x$ in eq.~(\ref{x}) numerically
the first consistency shows excellent agreement with the numerical 
results [like in the cases (a) and (c) in Fig.~\ref{suconsist}], 
as long as the correlation is not large at horizon crossing.

When the $\chi$ mass is light 
($|\delta|~\lsim~1$) and the second phase of inflation takes place, we 
found that the correlation $r_C$ is close to one, even changing the values 
of $g^2/\lambda$ to of order unity.  The correlation is also expected to 
be strong in other models of double inflation with a  
tachyonic instability.

\section{Conclusions}

In this paper we have studied the correlation of adiabatic and isocurvature
perturbations generated in inflationary scenarios with two phases of 
inflation (double inflation).
We have made a detailed multi-parameter numerical analysis of the 
power spectra relevant for the cosmic microwave background and large-scale 
structure.  We also studied the validity of the inflationary consistency 
relations derived from slow-roll analysis for two different models of the 
double inflation -- the noninteracting/interacting two massive scalar 
fields and the supersymmetric model with a tachyonic (spinodal) 
instability separating the two phases of inflation.

In single-field inflationary scenarios, the slow-roll approximation is 
typically reliable apart except near the end of 
inflation. In the case of multiple scalar fields, however, we 
need to be more careful in the use of the slow-roll approximation.  
If one of the scalar field is quickly suppressed and another scalar field 
leads to inflation with more than 60 e-folds, 
perturbations relevant for large-scale structure are effectively described 
by the single-field inflationary scenario.  
However, when both scalar fields are of the same order around 60 
e-folds before the end of double inflation, we are faced with limitations 
in the use of slow-roll results.  In this case the slow-roll parameter of a 
heavy scalar field is already large around the end of the first stage of 
inflation.

The assumption of the slow variation of the effective masses of 
``adiabatic" and ``entropy" fields, which is used to obtain the spectra of 
perturbations analytically, is often not valid in the context of double 
inflationary scenarios.  This is reflected in our results where we found 
that the slow-roll derived correlation $r_C$ and three spectral indices 
$n_{\cal R}$, $n_S$, $n_C$ do not agree well with the full numerical 
simulations, especially when the correlation is strong.  If the correlation 
is negligibly small {\it at horizon crossing}, the first consistency 
relation (\ref{consistency1}) shows good agreement with our numerical 
results [see the cases (a) and (b) in Fig.~\ref{consist} and the cases (a) 
and (c) in Fig.~\ref{suconsist}].  This is consistent with the result of 
Wands {\em et al.} that the first consistency relation was obtained by only 
assuming a vanishingly small correlation at horizon crossing 
\cite{Wands:2002bn}.  In the case where slow-roll conditions are violated 
at horizon crossing, which can occur in double inflationary scenarios, we 
find that numerical results exhibit some deviation from the first 
consistency relation (\ref{consistency1}) [see the case (c) in 
Fig.~\ref{consist} and the case (b) in Fig.~\ref{suconsist}].  

The second 
consistency relation (\ref{consistency2}) is more strongly affected by the 
change of the entropy/adiabatic mass and the scalar field velocity angle 
$\dot{\theta}$ during double inflation, thereby showing stronger 
deviations especially when the correlation is large.  These suggest the 
necessity of numerical analysis -- or a refined analytical treatment -- in 
order to correctly estimate the final power spectra, spectral indices and 
the correlations of perturbations.

We have also found that a wide variety of power spectra and correlations
can be obtained, depending on the parameters of the models considered.  In 
the case of noninteracting massive scalar fields, two important quantities 
determine the strength of the correlation: the ratio of the two scalar 
fields ($\tan \alpha_*$) and the ratio of the two masses ($R$).  We made a 
complete classification for several different cases to understand the 
correlation appropriately.  

When the interaction between two scalar fields 
($g^2\phi^2\chi^2$) is introduced, this can lead to a blue spectrum of 
isocurvature perturbations if the mass of the entropy field perturbation 
is 
larger than the Hubble rate.  However, the heavy field $\chi$ is soon 
suppressed toward the potential valley at $\chi=0$, in which case the 
correlation between adiabatic and isocurvature perturbations is weak.

Therefore the spectrum 
of the adiabatic perturbation is typically slightly red-tilted as in the 
case with $g=0$.  In this model we also found an interesting parameter 
range where large values of $g$ and $\chi$ lead to the rather steep 
red-tilted spectra of strong correlated adiabatic and isocurvature 
perturbations toward large scales.  This comes from the negative mass of 
the entropy field perturbation with comparable values of two scalar fields.

In the double inflationary scenario motivated by supersymmetric
theories, the correlation is found to be very large ($r_C \simeq 1$).
This is associated with a tachyonic growth of the entropy field
perturbation during the second stage of the double inflation.
This strong correlation also yields the mixture of 
adiabatic and isocurvature perturbations after the symmetry breaking, 
thereby modifying the spectra of perturbations generated during the first 
stage of inflation.  We found that a variety of power spectra can be 
obtained by making use of this conversion mechanism.

In the original version of the hybrid inflation with a 
potential (\ref{hybrid}) \cite{Linde:1993cn}, the field $\chi$ is 
strongly suppressed because of its large effective mass before 
the symmetry breaking.  
Inflation ends by a rapid rolling of the field $\chi$ after the symmetry 
breaking at $\phi=\phi_c$.  Since the field $\chi$ has practically 
no homogeneous component at $\phi=\phi_c$, 
the decomposition of $\chi$ between the homogeneous field $\chi(t)$ 
and the perturbative part $\delta \chi({\bf x}, t)$ is not necessarily 
valid.
When $\chi$ is negligibly small at $\phi=\phi_c$, we need to 
go beyond the perturbation theory using the spatial distribution 
of the field $\chi({\bf x}, t)$ as in ref.~\cite{Felder:2000hj}.  

Note, however, that in the case 
of the double inflation the field $\chi$ is hardly suppressed for 
$\phi>\phi_c$ due to the light $\chi$ mass ($|\delta|~\lsim~1$).
Then we are free from the problem of the decomposition of $\chi$,
in which case our linear analysis can be reliable.  We also made
some simulations including the back-reaction effect of field 
fluctuations as the Hartree approximation and 
obtained similar results as found in this work.

In our work we analyzed two models of the double inflation
given by the potentials (\ref{V}) and (\ref{hybrid}).
Since these potentials include most of the basic properties of the double 
inflation, it should be fairly easy to extend our analysis to other double 
inflation models motivated by particle physics.\footnote{In some models
of two-field inflation considered as in 
refs.~\cite{Starobinsky:1994mh,Tsujikawa:2000wc,Starobinsky:2001xq}, the 
second stage of inflation is absent.  In this case the first consistency 
relation (\ref{consistency1}) is expected to be valid, while the second 
one 
(\ref{consistency2}) may be model-dependent \cite{Wands:2002bn}.}

It is really 
encouraging that double inflation models lead to strong correlations 
over wide ranges of their parameter spaces. This suggests that searches 
for correlations in the CMB  
may yield interesting information and constraints on such models and 
motivates the development of enhanced slow-roll approximations which can 
accurately predict the full numerical results.

\section*{Appendix A:~Numerical methods to evaluate power spectra and 
correlations}

Let us explain the general numerical method used to calculate power spectra and 
correlations in the context of multi-field inflation.  We treat $Q_{\sigma}$ 
and $\delta s$ as independent stochastic variabes for the modes deep inside 
the Hubble radius.  Then we have to do two numerical runs in order to 
evaluate $P_{\cal R}$, $P_S$ and $P_C$.  One run corresponds to the 
Bunch Davies vacuum state for $Q_{\sigma}$ and $\delta s=0$ for the entropy 
field perturbation, in which case we get the solutions, ${\cal R}={\cal 
R}_1$ and $S=S_1$.  Another corresponds to the Bunch Davies vacuum state 
for $\delta s$ and $Q_{\sigma}=0$ for the adiabatic field perturbation, in 
which case we have ${\cal R}={\cal R}_2$ and $S=S_2$.

Then each power spectrum can be expressed in terms of 
${\cal R}_1$, ${\cal R}_2$, $S_1$, and $S_2$, as
\begin{eqnarray}
P_{\cal R}&=& \frac{k^3}{2\pi^2} \left( |{\cal R}_1|^2
+|{\cal R}_2|^2 \right)\,, 
\label{sto1} \\
P_S &=& \frac{k^3}{2\pi^2} \left( |S_1|^2 + |S_2|^2 \right)\,, \\
P_C &=& \frac{k^3}{2\pi^2} \left| {\cal R}_1S_1 + 
{\cal R}_2S_2 \right|\,.
\label{sto3}
\end{eqnarray}
{}From this it is easy to show that the correaltion $r_C=P_C/\sqrt{P_{\cal 
R}P_S}$ ranges $r_C \le 1$.

If we run the numerical code only once by using the initial conditions 
where both $Q_{\sigma}$ and $\delta s$ are in the vacuum state, we then 
get ${\cal R}={\cal R}_1+{\cal R}_2$.  In this case the power spectrum of 
${\cal R}$ yields $P_{\cal R}= \frac{k^3}{2\pi^2} \left| 
{\cal R}_1+{\cal R}_2 \right|^2$, which is different from 
eq.~(\ref{sto1}). 
As long as the perturbations are stochastic random variables initially, 
it is required to adopt the method described in 
eqs.~(\ref{sto1})-(\ref{sto3}).

\section*{ACKNOWLEDGEMENTS}
We thank Nicola Bartolo, Christopher Gordon, Julien Lesgourgues, Alexei 
Starobinsky,  Jun'ichi Yokoyama, and particularly 
Carlo Ungarelli and David Wands, for useful discussions.  
S.T.  is also thankful for 
financial support from the JSPS (No.  04942).  The research of BB is 
supported under PPARC grant PPA/G/S/2000/00115.  D.P.  is thankful for 
financial support from the Monbukagakusho Young Foreign Researcher summer 
program and grateful to RESCEU for hospitality.  S.T.  is grateful to 
Stanislav Alexeyev and Alexey Toporensky for kind hospitality during his 
stay at the Sternberg Astronomical Institute, Moscow State University.  


\end{document}